\documentclass{emulateapj}

\def\kms{\ifmmode{\rm km\thinspace s^{-1}}\else km\thinspace s$^{-1}$\fi} 
\def\ms{\ifmmode{\rm m\thinspace s^{-1}}\else m\thinspace s$^{-1}$\fi} 

\def\gamcep{$\gamma$~Cep}

\def\today{\number\year\space \ifcase\month\or  January\or February\or
        March\or April\or May\or June\or July\or August\or
September\or
        October\or November\or December\fi\space \number\day}

\shortauthors{Torres}
\shorttitle{\gamcep}

\begin{document}

\title{The planet host star $\gamma$~Cephei: physical properties, the
binary orbit, and the mass of the substellar companion}


\author{Guillermo Torres}

\affil{Harvard-Smithsonian Center for Astrophysics, 60 Garden St.,
Cambridge, MA 02138}
\email{gtorres@cfa.harvard.edu}

\begin{abstract} 

The bright K1\,III--IV star \gamcep\ has been reported previously to
have a companion in a $\sim$2.5-yr orbit that is possibly substellar,
and also has a stellar companion at a larger separation that has never
been seen. Here we determine for the first time the three-dimensional
orbit of the stellar companion accounting also for the perturbation
from the closer object. We combine new and existing radial velocity
measurements (of both classical precision and high precision) with
intermediate astrometric data from the {\it Hipparcos\/} mission
(abscissa residuals) as well as ground-based positional observations
going back more than a century. The orbit of the secondary star is
eccentric ($e = 0.4085 \pm 0.0065$) and has a period $P = 66.8 \pm
1.4$~yr and a semimajor axis of $19.02 \pm 0.64$~AU. We establish the
primary star to be on the first ascent of the giant branch, and to
have a mass of $1.18 \pm 0.11$~M$_{\sun}$, an effective temperature of
$4800 \pm 100$~K, and an age around 6.6~Gyr (for an assumed
metallicity [Fe/H] = $+0.01 \pm 0.05$). The unseen secondary star is
found to be an M4 dwarf with a mass of $0.362 \pm 0.022$~M$_{\sun}$,
and is expected to be $\sim$8.4~mag fainter than the primary in $V$
and $\sim$6.4~mag fainter in $K$. The minimum mass of the putative
planetary companion is $M_p \sin i = 1.43 \pm 0.13$~M$_{\rm Jup}$, the
inclination angle of its orbit being unknown.  Taking advantage again
of the high-precision {\it Hipparcos\/} observations we are able to
place a dynamical upper limit on this mass of 13.3~M$_{\rm Jup}$ at
the 95\% confidence level, and 16.9~M$_{\rm Jup}$ at the 99.73\%
(3$\sigma$) confidence level, thus confirming that it is indeed
substellar in nature. The orbit of this object (semimajor axis $1.94
\pm 0.06$~AU) is only 9.8 times smaller than the orbit of the
secondary star (the smallest ratio among exoplanet host stars in
multiple systems), but it is stable if coplanar with the binary.
	
\end{abstract}

\keywords{binaries: spectroscopic --- binaries: visual --- planetary
systems --- stars: individual (\gamcep) --- stars: late-type}

\section{Introduction}
\label{sec:introduction}

The bright evolved star $\gamma$~Cephei ($V = 3.21$, SpT =
K1~III--IV, $\alpha = 23^{\rm h} 39^{\rm m} 21\fs01$, $\delta =
+77\arcdeg 37\arcmin 55\farcs2$, J2000; also known as HD~222404,
HR~8974, HIP~116727) is among the first objects to be subjected to
high-precision radial-velocity measurements in an effort to discover
substellar-mass companions around nearby stars
\citep{Campbell:88}. This group of investigators \citep{Campbell:79}
pioneered the use of a hydrogen fluoride gas absorption cell on the
Canada-France-Hawaii telescope (CFHT) and achieved internal errors
around 13~\ms\ for bright stars, inaugurating the era of Doppler
searches that has been so successful in finding extrasolar planets in
the last 10 years.

Small radial-velocity variations in \gamcep\ were indeed seen by
\cite{Campbell:88} suggesting the presence of a Jupiter-mass object in
a $\sim$2.5-yr orbit. Those variations with a semi-amplitude of only
about 25~\ms\ were superimposed on a much larger variation caused by a
previously unnoticed stellar companion with a period of
decades. However, the interpretation of the residual 2.5-yr variation
as due to a planetary object was subsequently put in doubt by the same
group \citep[see][]{Irwin:89, Walker:89, Walker:92}. They argued that
changes with a similar period were observed in a chromospheric
activity indicator in \gamcep\ (the \ion{Ca}{2} $\lambda$8662 emission
line index), and thus that the velocity variations were spurious and
probably due only to changes in the spectral line profiles caused by
surface inhomogeneities (spots) driven by stellar rotation. More
recently the planetary interpretation was reinstated by
\cite{Hatzes:03} on the basis of new high-precision velocity
observations at the McDonald Observatory.  They showed convincingly
that the 2.5-yr variation is coherent in phase and amplitude
throughout the entire 20-yr interval covered by the merged CFHT and
McDonald data sets, as would be expected for Keplerian motion, and
that no changes were observed in the spectral line bisectors. On the
other hand, a careful re-analysis of the changes in the activity
indicator reported by \cite{Walker:92} revealed that the periodicity
of the \ion{Ca}{2} $\lambda$8662 measurements (2.14 yr) is not only
slightly different from that in the velocities, but it is transitory
in nature, thus ruling out a connection.

\gamcep\ also carries the distinction of being among the first planet
host stars to be found in a binary system, which raises interesting
issues related to the dynamical stability of such configurations. A
recent study by \cite{Raghavan:06} points out that among the known
planet host stars \gamcep\ happens to be the system with the smallest
ratio ($\sim$11) between the size of the binary orbit and the
planetary orbit. However, the outer orbit is at present poorly known,
and the secondary star is presumably very faint and has never been
seen. Reported values for the binary period have ranged between
29.9~yr \citep{Walker:92} and 66~yr \citep{Griffin:02}, and have been
based on only part of the data available, in some cases spanning much
less than a full cycle. A number of authors have carried out numerical
investigations of the gravitational influence of the secondary star on
the orbit of the planet \citep[e.g.,][]{Dvorak:03, Thebault:04,
Haghighipour:06}, but have used rather different parameters for the
binary or have pointed out the uncertainty in those elements as a
limiting factor.

The motivation for this paper is thus threefold: $i$) To improve the
determination of the orbit of the secondary star (including for the
first time an estimate of the inclination angle, and of the mass of
the secondary) in order to allow more definitive dynamical studies of
the stability and evolution of the system. We do this by using all
available radial velocity data for \gamcep\ including new measurements
reported here and other historical observations not previously
used. We incorporate also astrometric measurements from the {\it
Hipparcos\/} mission \citep[``abscissa residuals'';][]{ESA:97} as well
as transit circle and other positional information spanning more than
a century. $ii$) To carry out a critical review of previous studies of
the physical properties of the primary star and use all available
information to estimate its absolute mass, a key parameter influencing
the mass of the substellar companion. $iii$) To place firm dynamical
upper limits on the mass of this companion by taking advantage of the
high-precision {\it Hipparcos\/} intermediate data and modeling the
reflex motion of the primary star on the plane of the sky. We show
that this modeling allows us to confirm the substellar nature of the
companion, although it is not yet possible to rule out a mass in the
brown dwarf regime.

\section{Observational material}
\label{sec:observations}

We describe here all spectroscopic and astrometric measurements of
\gamcep\ of which we are aware that have a bearing on the motion of
the star, with the goal of combining them into a global orbital
solution in \S\ref{sec:orbit}.

\subsection{Radial velocities}
\label{sec:velocities}

The high-precision Doppler measurements of \gamcep\ have been
described in detail by \cite{Hatzes:03}. They consist of 4 separate
data sets corresponding to different instrument configurations:
three from the McDonald Observatory \citep[referred to below as
McDonald~I, McDonald~II, and McDonald~III, following][]{Hatzes:03},
and one from the CFHT, which is the data set of \cite{Walker:92}. The
nominal precision of these measurements ranges from about 8~\ms\ to
$\sim$30~\ms, and they are all differential in nature as they rely on
the use of telluric O$_2$ lines as the velocity metric, or on lines of
hydrogen fluoride or iodine gas that play the same role. Taken
together these velocities cover the interval 1981.4--2002.9, which
includes periastron passage in the binary orbit. \cite{Hatzes:03}
combined these data and solved simultaneously for the outer orbit and
the orbit of the planet. The time span of the observations is less
than half of their estimated binary period of 57~yr.

Beginning in the late 1970's \gamcep\ was monitored spectroscopically
using more traditional means by \cite{Griffin:02}. To their own
observations with several different instruments in Cambridge
(England), Haute-Provence (France), and Victoria (Canada), they added
a subset of the high-precision velocities mentioned above as well as
other velocities collected from the literature in an effort to extend
the time coverage and better constrain the outer orbit. These include
measures published by \cite{Beavers:86} made in 1978--1980, and most
importantly the velocities obtained at the Lick Observatory in
1896--1921 \citep{Campbell:28}. All these measurements were placed by
\cite{Griffin:02} on the same zero point (corresponding to their
Cambridge instrument), and have formal uncertainties ranging from
0.2~\kms\ to 0.9~\kms.  We adopt these 77 measurements as published.
These authors noted an unfortunate gap of some 50 years in the
velocity coverage for \gamcep\ that complicates the determination of
the orbital period (see also \S\ref{sec:period}). In order to
distinguish between two possible periods (66~yr and 77~yr) allowed by
the radial velocity data they used, they considered also other
measurements from the literature in the interval 1902--1907
\citep{Frost:03, Belopolsky:04, Slipher:05, Kustner:08}. They found
that those velocities favored the 66-yr period, although they did not
actually make use of them in their orbital solution because of their
uncertain zero point.

Our own contribution to the observational material is twofold. On the
one hand we have derived 3 new velocities for \gamcep\ based on
archival spectra collected at the Harvard-Smithsonian Center for
Astrophysics (CfA) using an echelle spectrograph on the 1.5-m
Tillinghast reflector at the F.\ L.\ Whipple Observatory. The nominal
precision of these measurements is around 0.3~\kms\ for a bright and
sharp-lined star such as this. For details on the reduction procedures
we refer the reader to the description by \cite{Torres:02}. While this
contribution is modest by comparison to the material described
earlier, it does extend the time coverage to the end of 2004, and the
velocities are on a well-defined system \citep[see][]{Stefanik:99,
Latham:02}.

Given the poor spectroscopic coverage prior to 1978, we carried out a
careful search of the literature for additional measurements that
might help constrain the outer orbit. Aside from the 1902--1907
sources mentioned above that were used only as supporting evidence by
\cite{Griffin:02}, a number of other velocity sources were found but
their zero points are generally unknown, so the measurements cannot be
combined at face value. Thus as a second contribution we relied on the
extensive CfA database of $\sim$$250,\!000$ spectra to place all of
those scattered measurements of \gamcep\ onto a uniform frame of
reference. This was accomplished by using measurements for other stars
also reported in each of these sources, and comparing them with newly
derived velocities for those same ``standards'' from CfA spectra
obtained at one time or another over the past 25 years.  Details of
this procedure are provided in the Appendix. Of particular relevance
are the \gamcep\ measurements by \cite{Kjaergaard:81} made in 1977,
\cite{Snowden:05} in 1972--1974, \cite{Boulon:57} in 1955, and
\cite{Harper:34} in 1921.  The precision of those velocities ranges
from 0.5~\kms\ to about 1.9~\kms. We list them in
Table~\ref{tab:rvcfa} along with our own measurements, all on the CfA
system.

Despite our attempts to establish their zero points, two of the
sources of historical velocities showed large discrepancies when
compared with other data taken at similar times, or presented other
problems.  The series of measurements by \cite{Belopolsky:04} contains
only 5 other stars usable as standards, and the offset required to
place those velocities on the CfA system has the largest uncertainty
($\sim$1~\kms). The corrected \gamcep\ velocities from 1903 are some
3~\kms\ too high. Three velocities measured at Mt.\ Wilson Observatory
in 1915--1917 \citep{Abt:73} show the largest spread of any data set
(4~\kms). The zero point of those measurements is very difficult to
establish because of the variety of instruments and telescopes used,
which are not always indicated in the original publication. The
average of the corrected velocities for \gamcep\ shows a discrepancy
of 5~\kms\ relative to others made within a few years. We have
therefore not made use of either of these two data sets in our orbital
solution described in \S\ref{sec:orbit}. Several high-dispersion
plates of \gamcep\ were obtained by \cite{Koelbloed:75} in 1963--1964,
a critical time in the observational history of this object, but
unfortunately the authors appear not to have measured radial
velocities.  Finally, \cite{Rucinski:81} published a velocity
measurement for \gamcep\ made in 1979, but all of the other stars
reported in their paper happen to be variable, and so cannot be used
as standards.

\subsection{Astrometry}
\label{sec:astrometry}

\gamcep\ was observed by the {\it Hipparcos\/} satellite between 1989
and 1993 \citep{ESA:97}. These accurate one-dimensional astrometric
measurements were used by the Science Team to derive the position,
proper motion, and trigonometric parallax of the object ($\pi_{\rm
HIP} = 72.50 \pm 0.52$~mas) as reported in the main catalog. The
astrometric solution revealed a measurable acceleration on the plane
of the sky (proper motion derivatives) in the amount of
$d\mu_{\alpha}/dt = +1.51 \pm 1.12$~mas~yr$^{-2}$ in Right Ascension
and a more significant $d\mu_{\delta}/dt = +6.10 \pm
1.11$~mas~yr$^{-2}$ in Declination. This acceleration is of course due
to the binary nature of the object, and was accounted for in deriving
the parallax.

As we demonstrate below, the binary motion at the epoch of the {\it
Hipparcos\/} observations is such that we expect some curvature on the
plane of the sky that should be detectable in the measurements. We
have therefore made use of these observations (available in the form
of ``abscissa residuals'') in our orbital solution described below,
since they are complementary to the spectroscopic observations and
provide new information. A total of 76 such measurements were obtained
by the two independent data reduction consortia \citep{ESA:97}, and
the median error for a single measurement is 1.9~mas.

Because of the relatively short time span of these observations
compared to the binary orbital period, it is almost certain that part
of the orbital motion has been absorbed into the proper motion
components reported by {\it Hipparcos\/}. This is in fact a way in
which many long-period binaries have been discovered in the past, on
the basis of the apparent variability of their proper motions when
computed at different epochs \citep[see, e.g.,][]{Wielen:99,
Gontcharov:00, Makarov:05}. Precisely this effect was pointed out for
\gamcep\ by \cite{Heintz:90}, who noticed a significant change mostly
in $\mu_{\delta}$ over several decades. Therefore, to make proper use
of the {\it Hipparcos\/} intermediate data to extract information on
the binary orbit it is necessary to constrain the proper motion by
other means in order to model the orbital motion without risking
systematic errors.

Initially we considered using the proper motion for \gamcep\ reported
in the Tycho-2 catalog \citep{Hog:00a}, which relies on ground-based
positional measurements made over many decades and is constrained at
the recent epoch by the Tycho-2 position. This long baseline
presumably averages out any perturbations due to orbital motion if the
period is significantly shorter than this. The Tycho-2 proper motion
is in fact quite different from the {\it Hipparcos\/} determination,
which is effectively ``instantaneous'' at the mean epoch
$\sim$1991.25. In the case of \gamcep, however, the orbital period is
not negligible compared to the time span, and we were concerned that
$\mu_{\alpha}$ and $\mu_{\delta}$ might be biased. Evidence that the
orbital motion is detectable in the individual positional measurements
from transit circle observations was indeed presented by
\cite{Gontcharov:00}, who inferred from them a period of about 45~yr
for the binary.  We therefore chose to make use of the individual
positions from ground-based catalogs going back to 1898, kindly
provided by S.\ Urban of the U.S.\ Naval Observatory (USNO).
Additional measurements from the Carlsberg Meridian Catalogs
\citep[CMC; see, e.g.,][]{Carlsberg:89} were provided by G.\
Gontcharov (Pulkovo Observatory) or obtained from the literature. All
of these measurements have been reduced to the International Celestial
Reference Frame (ICRF), effectively represented in the optical by the
{\it Hipparcos\/} catalog, and their nominal precision varies between
about 50 and 500~mas \citep[see][]{Hog:00b}. We list them in
Table~\ref{tab:astrometry}.
	
\section{Orbital solution}
\label{sec:orbit}

The combination of the radial velocity mesurements and the astrometry
makes it possible to derive the complete set of elements describing
the binary orbit in \gamcep. The inclination angle is of particular
interest because when combined with the spectroscopic mass function it
provides the information needed to compute the mass of the secondary
star, given an estimate of the primary mass. The substellar companion
to the primary introduces additional components of motion that we
model simultaneously. Given that the outer orbit is an order of
magnitude larger than the inner orbit (see \S\ref{sec:introduction}),
to first order we assume here that they are decoupled, i.e., that the
outer one may be treated as corresponding to a ``binary'' composed of
the secondary star (B) and the center of mass of the inner pair (A).
Orbital elements that refer to the outer orbit are indicated below
with the subindex ``AB'', and those pertaining to the inner orbit are
distinguished with a subindex ``A''.  The primary star itself is
referred to as ``Aa'' following the traditional spectroscopic
notation, and the planet (indistinctly called also `substellar
companion') as ``$p$'', for simplicity.

The radial velocities allow us to solve for the period, center-of-mass
velocity of the (triple) system, eccentricity, velocity
semi-amplitude, longitude of periastron, and time of periastron
passage in the outer orbit: \{$P_{\rm AB}$, $\gamma$, $e_{\rm AB}$,
$K_{\rm A}$, $\omega_{\rm A}$, $T_{\rm AB}$\}. The high-precision
velocities constrain the spectroscopic elements of the inner
(planetary) orbit: \{$P_{\rm A}$, $e_{\rm A}$, $K_{\rm Aa}$,
$\omega_{\rm Aa}$, $T_{\rm A}$\}. Because the high-precision
velocities are differential, an offset must be determined to place
them on the frame of the absolute velocities, for which we have chosen
the Griffin data set as the reference. We therefore solved for 4
additional parameters representing these offsets, one for each data
set: $\Delta RV_1$, $\Delta RV_2$, and $\Delta RV_3$ for the groups
referred to as McDonald~I, McDonald~II, and McDonald~III (see
\S\ref{sec:velocities}), and $\Delta RV_4$ for the CFHT data set. The
CfA velocities and other historical data sets placed on the CfA system
were considered as a single group, and one additional parameter
$\Delta RV_5$ was included to represent the shift relative to Griffin.

Preliminary estimates suggested the secondary star is very small
compared to the primary, and we may assume here that it contributes no
light.  The {\it Hipparcos\/} observations therefore refer strictly to
the primary as opposed to the center of light, and provide a
constraint on the orientation of the outer orbit (inclination angle
$i_{\rm AB}$ and position angle of the ascending node $\Omega_{\rm
AB}$, referred to the equinox of J2000) as well as on the angular
scale of the orbit of the inner binary relative to the barycenter
($a''_{\rm A}$). We show below that the astrometric measurements do
not, however, resolve the wobble of the primary star caused by the
planet. We point out also that there are no available measurements of
the relative position between the two stars, since the secondary has
never been resolved. The use of the {\it Hipparcos\/} measurements in
the global solution introduces several other parameters that must be
solved for, including corrections to the catalog values of the
position of the barycenter ($\Delta\alpha^*$, $\Delta\delta$) at the
mean reference epoch of 1991.25, and corrections to the proper motion
components ($\Delta\mu_{\alpha}^*$,
$\Delta\mu_{\delta}$)\footnote{Following the practice in the {\it
Hipparcos\/} catalog we define $\Delta\alpha^* \equiv \Delta\alpha
\cos\delta$ and $\Delta\mu_{\alpha}^* \equiv \Delta\mu_{\alpha}
\cos\delta$.}. In principle we also need to solve for a correction to
the {\it Hipparcos\/} parallax. However, the fact that the
spectroscopic elements of the outer orbit are solved for at the same
time introduces a redundancy, and the parallax (which in this case
would be termed an ``orbital'' parallax) can be expressed in terms of
other elements as
\begin{equation}
\pi = 1.0879 \times 10^4 {a''_{\rm A} \sin i_{\rm AB} \over P_{\rm
AB}~K_{\rm A} \sqrt{1 - e_{\rm AB}^2}}~,
\label{eq:eq1}
\end{equation}
where the period is given in days and $K_{\rm A}$ in \kms. We have
therefore chosen to eliminate the parallax correction as an adjustable
parameter.

By combining complementary observations of different kinds the global
solution is strengthened.  The ground-based positional measurements
provide the tightest constraint on the proper motion and position of
the barycenter. This breaks the strong correlation between proper
motion and orbital motion in the {\it Hipparcos\/} observations, and
enables those measurements to provide a constraint on the angular
scale and orientation of the outer orbit, even though their coverage
is only a small fraction of the orbital period.  Some information on
the scale and orientation, as well as on the outer period, is provided
also by the positional measurements, while the velocities contribute
most of the weight to the period, shape, and linear scale of the
binary orbit. The elements of the inner orbit are constrained only by
the high-precision velocity measurements, and are only weakly
dependent on the outer orbit. Light-travel effects in the inner orbit
are negligibly small. The formalism for incorporating the abscissa
residuals from {\it Hipparcos\/} into the fit follows closely that
described by \cite{vanLeeuwen:98} and \cite{Pourbaix:00}, including
the correlations between measurements from the two independent data
reduction consortia \citep{ESA:97}. In using the ground-based catalog
positions the parallactic motion was accounted for in our model, given
that the precision of some of the more recent measurements is
comparable to the parallax.

Altogether there are 23 unknowns that we solved for simultaneously,
using standard non-linear least-squares techniques \citep[][p.\
650]{Press:92}. The solution converged quickly from initial values of
the elements chosen from preliminary fits or by an extensive grid
search, and experiments in which we varied the initial values within
reason yielded the same results.  A total of 446 individual
observations were used from 10 different data sets, as follows: 107
classical radial velocities (77 from Griffin, 30 from CfA and other
literature sources), 199 high-precision velocities (68 from CFHT, 43
from McDonald~I, 49 from McDonald~II, and 39 from McDonald~III), 76
one-dimensional {\it Hipparcos\/} measurements, and 64 ground-based
catalog coordinates (split into two data sets of 15 and 17 pairs of
Right Ascension and Declination measurements).  Weights were assigned
to the measurements according to their individual errors. Since
internal errors are not always realistic, we adjusted them by applying
a scale factor in such a way as to achieve a reduced $\chi^2$ value
near unity separately for each data set.  This was done by
iterations. These scale factors were all close to unity for most of
the velocity sets, and somewhat larger for some of the ground-based
catalog positions\footnote{The scale factors derived are 0.96 (Griffin
velocities), 0.93 (McDonald~I), 1.10 (McDonald~II), 1.02
(McDonald~III), 1.45 (CFHT), 0.94 (CfA), 1.62 (USNO Right Ascensions),
1.50 (USNO Declinations), 0.83 (CMC Right Ascensions), 0.87 (CMC
Declinations), and 0.86 ({\it Hipparcos\/}).}.

The results are given in Table~\ref{tab:elements}, along with derived
quantities such as the position of the barycenter at the mean epoch of
the {\it Hipparcos\/} catalog (1991.25), the parallax and proper
motion components, and the mass function of the stellar binary. Other
derived quantities are described below. The elements of the planetary
orbit are not significantly different from those reported by
\cite{Hatzes:03} since that orbit depends essentially only on the
high-precision velocities, for which we used the same data they used.
The parallax is also not appreciably different from the {\it
Hipparcos\/} value, although our uncertainty is somewhat smaller. The
proper motion components, on the other hand, are considerably
different from their catalog values, as anticipated above.

\begin{figure}
\vskip 0.4in
\epsscale{0.95}
\plotone{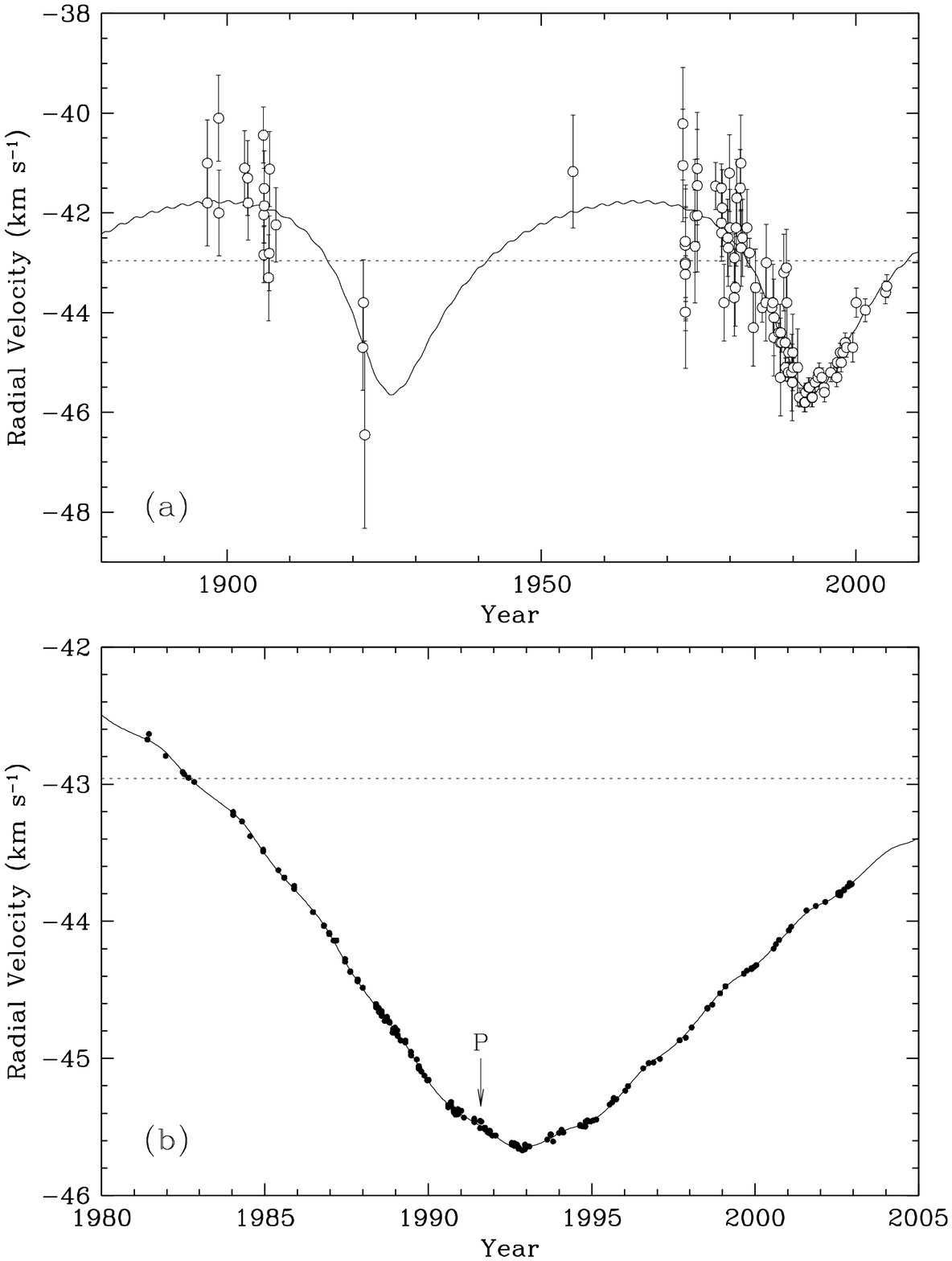}
\vskip -0.3in

\figcaption[]{Radial velocity measurements of \gamcep~A as a function
of time, along with our fitted curve from the combined solution. The
center-of-mass velocity of the system is indicated by the dotted line,
and is on the reference frame of the velocities by
\cite{Griffin:02}. (a) Classical velocity measurements in the outer
orbit. The wiggles in the curve correspond to the perturbation by the
2.47-yr substellar companion. (b) Close-up of the high-precision
velocities, which are near periastron passage in the outer orbit
(arrow). The error bars in this panel are smaller than the size of the
points. \label{fig:rvs1}}

\end{figure}

\begin{figure}
\vskip 0.5in
\epsscale{0.95}
\plotone{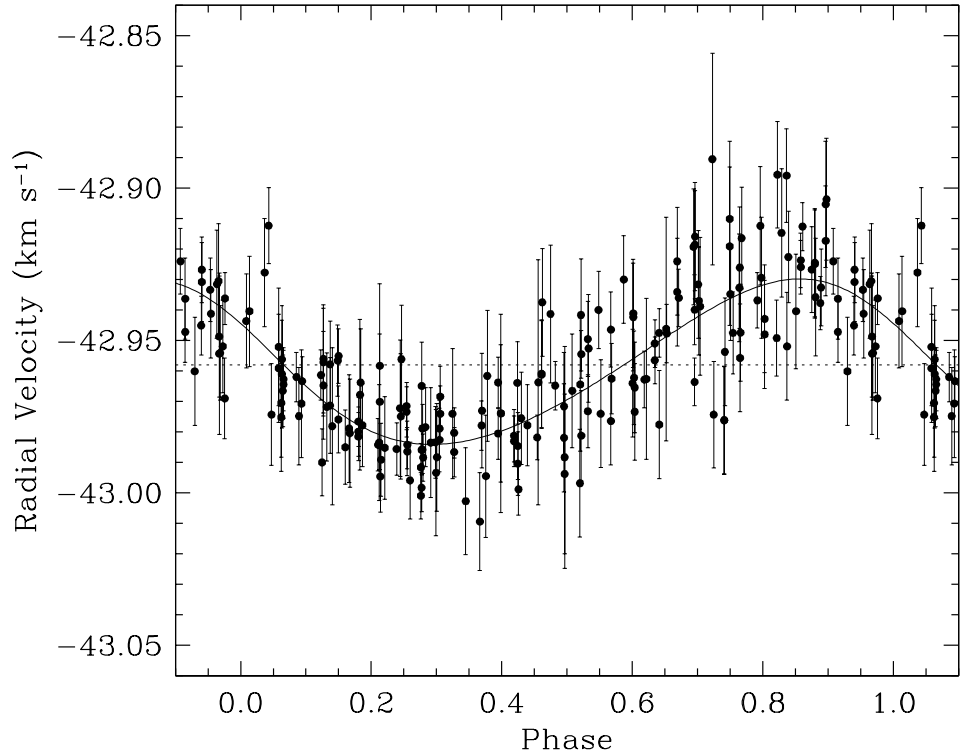}
\vskip -1.8in

\figcaption[]{High-precision radial velocity measurements of \gamcep\
 shown as a function of phase in the inner orbit, along with our
 fitted curve from the combined solution. The motion in the outer
 orbit has been subtracted. The center-of-mass velocity of the system
 is indicated by the dotted line, and is on the reference frame of the
 velocities by \cite{Griffin:02}. \label{fig:rvs2}} 

\end{figure}

The radial velocity measurements are represented graphically in
Figure~\ref{fig:rvs1}. The top panel shows only the classical
measurements. The small undulations in the computed curve are produced
by the wobble of the primary star with the 2.47-yr period of the
substellar companion.  Although the phase coverage in the outer orbit
is incomplete, the period of the binary is fairly well established
thanks in part to the very high precision of the McDonald and CFHT
data (see also \S\ref{sec:period}). These are shown separately in the
lower panel, where the error bars are smaller than the size of the
points. The reflex motion of the primary due to the substellar
companion is shown as a function of orbital phase in
Figure~\ref{fig:rvs2}, where the motion in the outer orbit has been
subtracted from the individual data sets. The residuals from the new
CfA velocities and from other values from the literature sources
placed on the same system are given in Table~\ref{tab:rvcfa}.

\begin{figure}
\vskip 0.3in
\epsscale{1.00}
\plotone{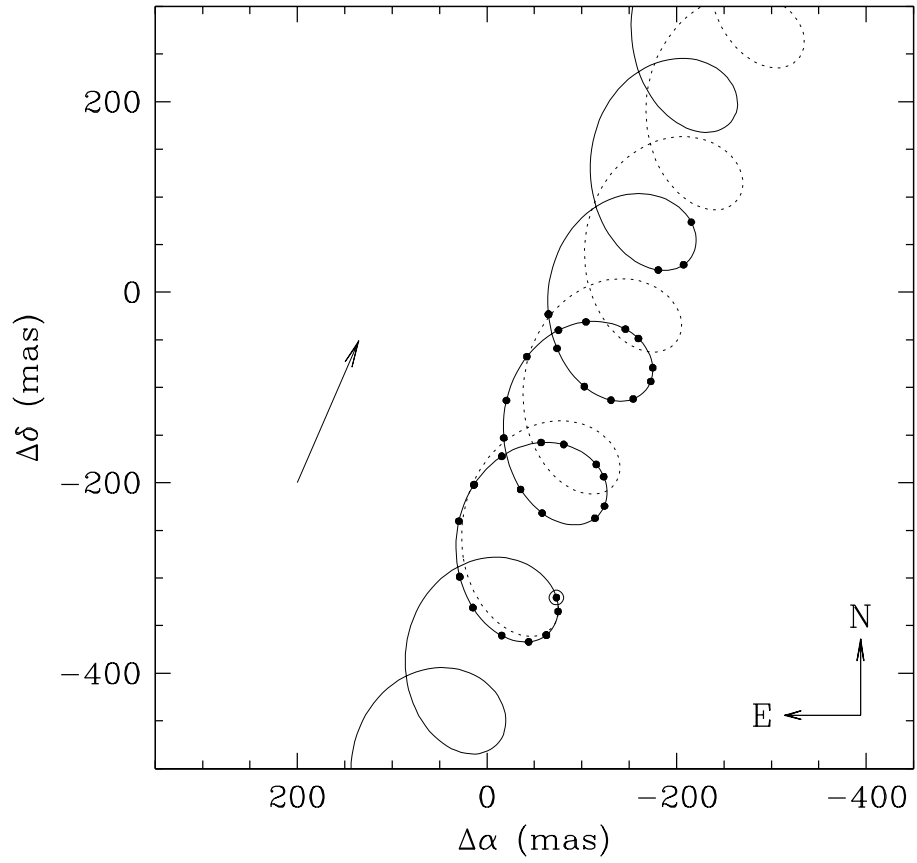}
\vskip -1.3in

\figcaption[]{Path of \gamcep~A on the sky resulting from the combined
 effects of proper motion, orbital motion ($P = 66.8$~yr), and
 parallactic motion (solid curve). The magnitude and direction of the
 annual proper motion is indicated by the arrow. The {\it Hipparcos\/}
 observations are shown as dots at their predicted locations, and do
 not represent the actual measurements, which are one-dimensional in
 nature (see text). The dotted curve shows the path the star would
 follow in the absence of orbital motion, starting at the epoch of the
 first {\it Hipparcos\/} measurement, indicated with the open
 circle. \label{fig:hiporb3}}

\end{figure}

The path of \gamcep\ on the plane of the sky is represented in
Figure~\ref{fig:hiporb3}, where the axes are parallel to the Right
Ascension and Declination directions. The solid curve is the result of
the contributions from the annual proper motion (arrow), the
parallactic motion, and the motion in the binary.  The wobble due to
the substellar companion is negligible on the scale of this figure
(see also \S\ref{sec:planet}). The predicted location of the
one-dimensional {\it Hipparcos\/} observations is indicated by the
dots on the curve. As stated earlier, their typical uncertainty is 1.9
mas. For illustration purposes, the dotted line in the figure starting
at the location of the first {\it Hipparcos\/} observation shows the
path the star would follow \emph{without} the perturbation from the
orbital motion in the binary.

\begin{figure}
\vskip 0.3in
\epsscale{1.00}
\plotone{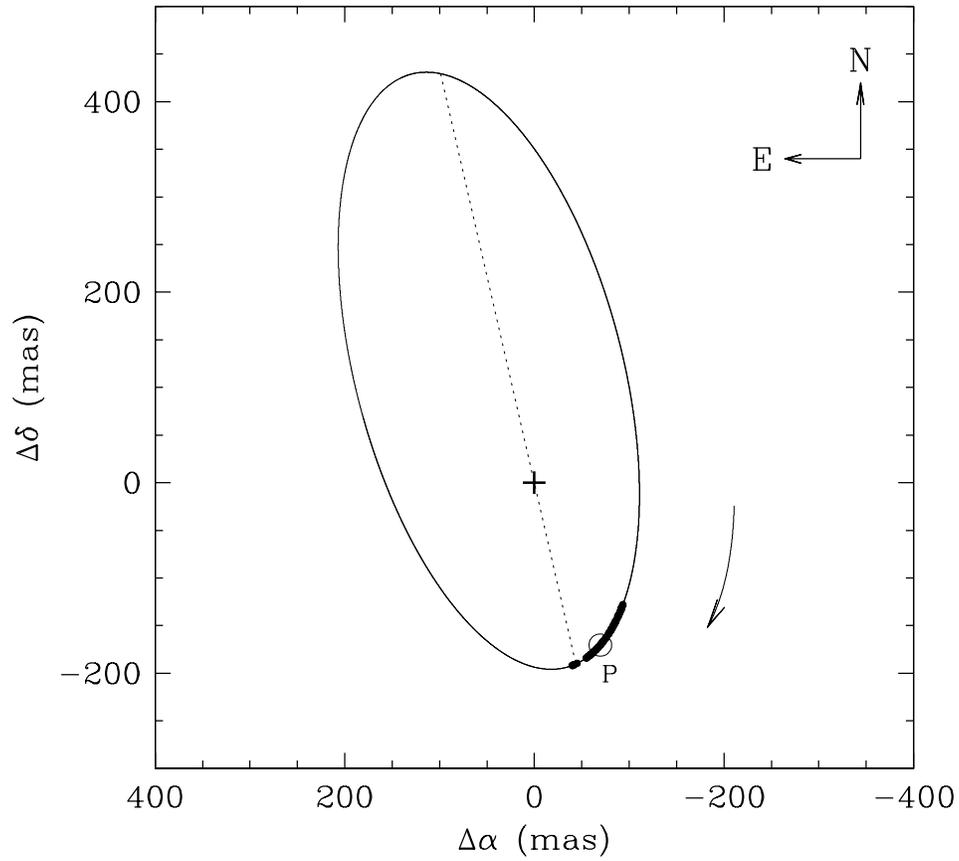}
\vskip -1.3in

\figcaption[]{Computed orbit of \gamcep~A around the center-of-mass of
 the binary (shown with a plus sign). The direction of motion
 (retrograde) is indicated by the arrow, and the dotted line
 represents the line of nodes. The {\it Hipparcos\/} observations are
 displayed with filled circles at their predicted locations, and are
 seen to bracket periastron passage (open circle labeled ``P''). The
 perturbation due to the substellar companion of the primary is
 negligible on the scale of this figure. The relative orbit of the
 binary is simply a scaled-up version of the ellipse shown here, with
 a scale factor given by $(M_{\rm Aa} + M_{\rm B})/M_{\rm B} = 4.26$
 (yielding a semimajor axis $a''_{\rm AB} = 1\farcs382 \pm
 0\farcs047$; see Table~\ref{tab:elements}).\label{fig:hiporb2}}

\end{figure}

\begin{figure}
\vskip 0.3in
\epsscale{1.00}
\plotone{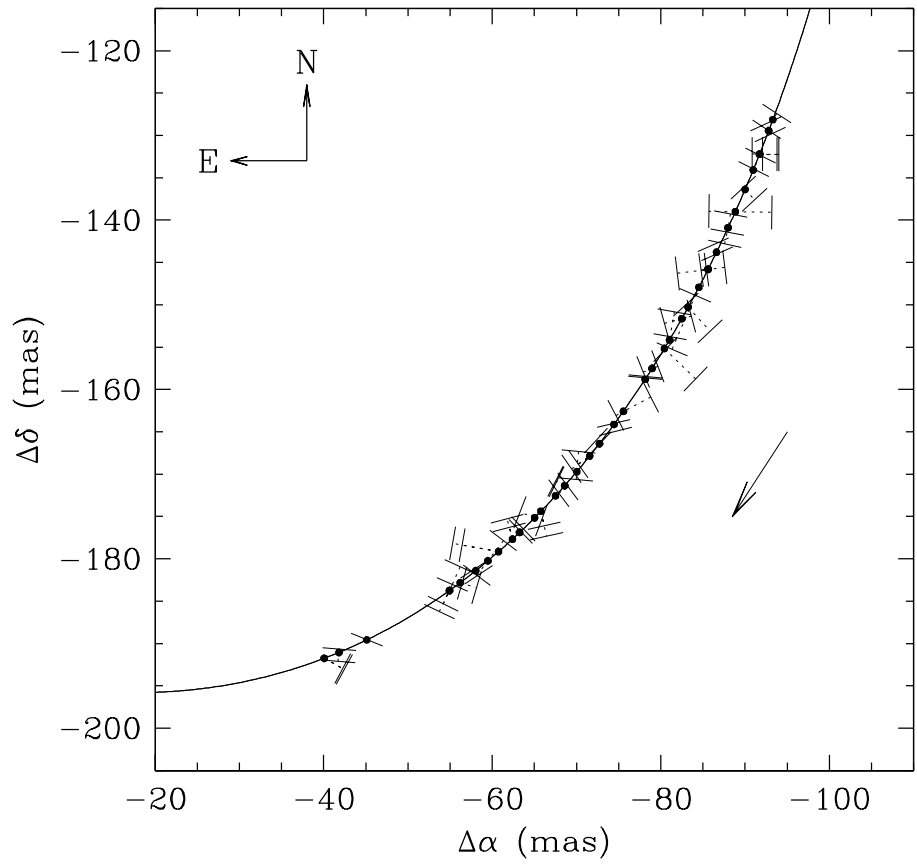}
\vskip -1.3in
 \figcaption[]{Enlargement of Figure~\ref{fig:hiporb2} showing the
 individual {\it Hipparcos\/} observations. See text for an
 explanation of the graphical representation of these one-dimensional
 measurements. \label{fig:hiporb2e}}
 \end{figure}

The orbit of the primary star in \gamcep\ around the center of mass of
the binary is shown in Figure~\ref{fig:hiporb2}, with a semimajor axis
of about 325 mas. The direction of motion is retrograde (arrow). The
intersection between the orbital plane and the plane of the sky (line
of nodes) is represented by the dotted line. The section of the orbit
covered by the {\it Hipparcos\/} mission is indicated with filled
circles, and the open circle labeled ``P'' represents periastron. A
close-up of the area around the {\it Hipparcos\/} observations is
shown in Figure~\ref{fig:hiporb2e}.  Because these measurements are
one-dimensional in nature, their exact location on the plane of the
sky cannot be shown graphically. The filled circles represent the
predicted location on the computed orbit. The dotted lines connecting
to each filled circle indicate the scanning direction of the {\it
Hipparcos\/} satellite for each measurement, and show which side of
the orbit the residual is on. The short line segments at the end of
and perpendicular to the dotted lines indicate the direction along
which the actual observation lies, although the precise location is
undetermined.  Occasionally more than one measurement was taken along
the same scanning direction, in which case two or more short line
segments appear on the same dotted lines. 

\begin{figure}
\vskip 0.2in
\epsscale{1.00}
\plotone{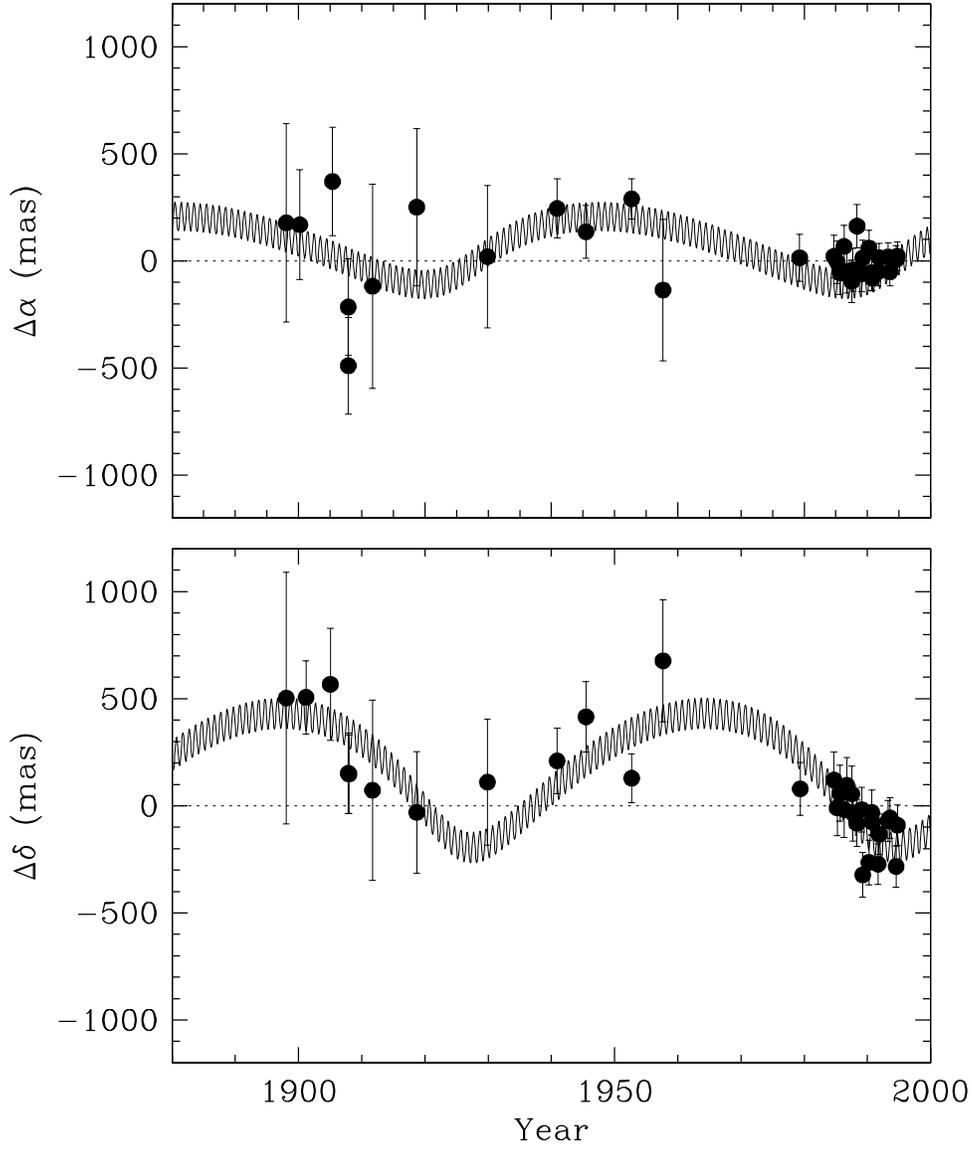}
\vskip -0.7in

 \figcaption[]{Ground-based catalog positions of \gamcep\ in Right
 Ascension and Declination, after subtracting the contribution from
 the proper motion resulting from our fit between the date of each
 observation and the reference epoch 1991.25. The curve represents the
 combination of motion in the 66.8-yr binary orbit and the parallactic
 motion, as predicted from the solution.\label{fig:wobble1}}

 \end{figure}

\begin{figure}
\vskip 0.2in
\epsscale{1.00}
\plotone{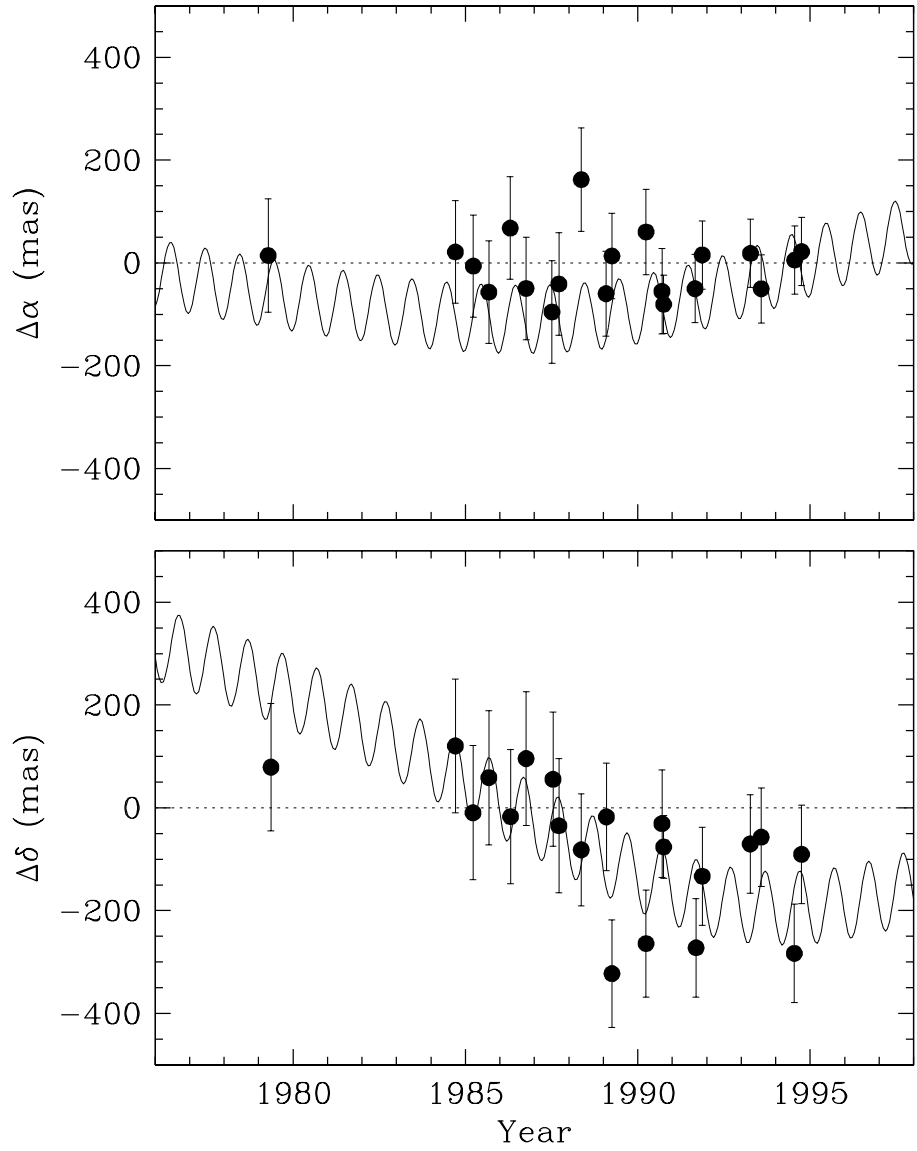}
\vskip -0.7in

 \figcaption[]{Enlargement of Figure~\ref{fig:wobble1} showing only
 the most recent catalog measurements, which are the most
 precise.\label{fig:wobble2}}

 \end{figure}

The motion of \gamcep~A in the binary orbit is discernible in the
ground-based catalog measurements taken over the last century,
although only in the Declination direction. This is illustrated in
Figure~\ref{fig:wobble1}. The amplitude of motion in the R.A.\
direction is much smaller because of the orientation of the orbit,
which is mostly North-South. The more recent measurements since 1980
are much more precise. That section of the orbit is shown on a larger
scale in Figure~\ref{fig:wobble2}. The residuals of all ground-based
measurements from our orbital solution are given in
Table~\ref{tab:astrometry}.

\subsection{The constraint on the binary period}
\label{sec:period}

The poor observational coverage of \gamcep\ prior to 1980 has made it
difficult to establish the period of the outer orbit in previous
studies, particularly since the secondary star has never been
resolved. \cite{Walker:92} gave a rough estimate of 29.9~yr based on
only 10~yr of high-precision velocity coverage. \cite{Gontcharov:00}
used the ground-based catalog positions spanning a little less than
six decades and inferred a period of 45~yr. A re-reduction of those
same data (G.\ Gontcharov 2006, private communication) does not show
that periodicity as clearly, however.  The radial-velocity study by
\cite{Griffin:02} took advantage of some of the historical
measurements going back more than 100~yr, and found that a 50-yr gap
in the data near the middle of the last century allowed two possible
periods giving fits of similar quality: $\sim$66~yr and
$\sim$77~yr. On the basis of other observational evidence they chose
the short orbital period.  \cite{Hatzes:03} used 20~yr worth of
high-precision velocity measurements and derived a period of 57~yr.

\begin{figure}
\vskip 0.4in
\epsscale{1.00}
\plotone{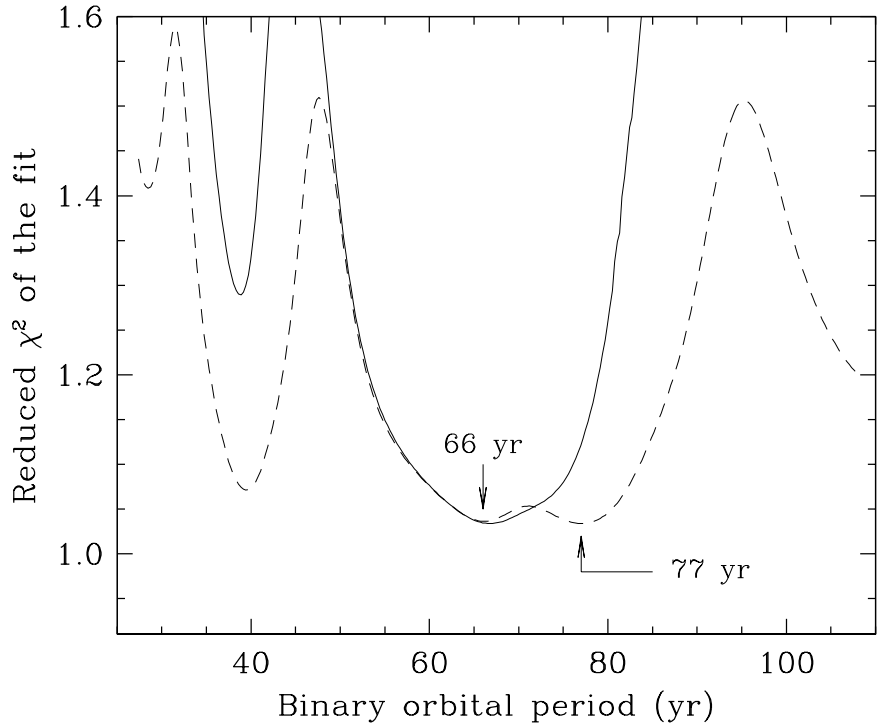}
\vskip -1.7in

 \figcaption[]{Constraint on the orbital period of the binary for two
 different data sets: the one in this paper (solid line), and the one
 used by \cite{Griffin:02} (dashed line). In each case the period has
 been fixed over a fine grid of values and the remaining elements were
 solved for in the usual manner. The run of the reduced $\chi^2$
 values for the present solution shows that of the two periods allowed
 by the \cite{Griffin:02} fit, the 66-yr value is the correct
 one. \label{fig:period}}

 \end{figure}

The simultaneous use of all of the above measurements, and the
addition of other observations (including more recent velocities as
well as historical velocities, and the {\it Hipparcos\/}
measurements), have allowed us to finally constrain the binary period
without ambiguity to a value of 66.8~yr, thus proving
\cite{Griffin:02} essentially correct. To illustrate the improvement
brought about by the added observations, we have recreated the fit by
\cite{Griffin:02} by using the same set of observations they
used\footnote{The only difference in our recreated solution is that we
used the high-precision velocity measurements as published, whereas
\cite{Griffin:02} used values read off from a figure, since some of
the measurements had not yet been reported in tabular form in the
literature.}, ignoring the velocity perturbation from the substellar
companion, as they did. In Figure~\ref{fig:period} we show the reduced
$\chi^2$ of the fit for a range of fixed orbital periods, with the
remaining orbital elements adjusted as usual to minimize $\chi^2$. We
then repeated this exercise using the data that went into our own
solution, this time accounting properly for the planetary
companion. The dashed curve corresponding to the solution by
\cite{Griffin:02} shows two local minima at $\sim$66~yr and
$\sim$77~yr, as found by those authors.  The solid curve corresponding
to the solution in this paper that includes all available observations
has a single minimum at 66.8~yr. The formal uncertainty in this value
is 1.4~yr, or 2\%.

\section{Physical properties of the primary and secondary stars}
\label{sec:properties}

In order to take full advantage of the orbital solution presented
above and estimate the mass of the unseen secondary star, as well as
to place limits on the mass of the substellar companion, we require an
estimate of the mass of the primary star itself ($M_{\rm Aa}$).  Mass
estimates in the literature for \gamcep\ have varied by more than a
factor of two (between $\sim$0.8~M$_{\sun}$ and $\sim$1.7~M$_{\sun}$),
which is somewhat surprising for such a bright and well-studied star
but may perhaps be explained by its present evolutionary state and
other uncertainties (see below). We wish to constrain it to much
better than this to avoid propagating the uncertainty to other
quantities that depend on the mass. In this section we therefore
examine the available observational material carefully and critically,
making use of current stellar evolution models to arrive at the best
possible estimate for $M_{\rm Aa}$. We discuss some of the other
estimates as well in an attempt to understand the differences.

The brightness of \gamcep\ has made it an easy target for
spectroscopic studies to determine both the effective temperature and
chemical composition of the star. These, along with other properties,
are essential in order to estimate its absolute mass.  In
Table~\ref{tab:specmet} we have collected the results of nearly two
dozen separate investigations carried out over the past 40 years.  We
consider here only determinations of $T_{\rm eff}$ and [Fe/H] that are
purely spectroscopic. Our own temperature estimate from the spectra
described in \S\ref{sec:velocities} is listed as well (see
\S\ref{sec:appendix} for the details of our procedures).  For the most
part the 22 independent metallicity determinations show reasonable
agreement within the errors, and yield a weighted average of [Fe/H] =
$+0.01 \pm 0.02$, or very nearly solar. Further comments on this value
are given below. The weighted average effective temperature is $T_{\rm
eff} = 4852 \pm 26$~K from 9 spectroscopic measurements including our
own.  The uncertainties given here are statistical errors that account
for the different weights as well as the scatter of the individual
[Fe/H] and $T_{\rm eff}$ measurements, but not for possible
systematics. In the following we adopt for these averages more
conservative errors of 0.05 dex and 100~K, respectively.

Temperature estimates for the star have been derived on numerous
occasions also from color indices in a variety of photometric systems.
In order to bring homogeneity to this information we have compiled the
available photometry for \gamcep\ in eight different systems (Johnson,
Str\"omgren, Vilnius, Geneva, Cousins, DDO, 2MASS, and Tycho) mostly
from the photometric database maintained by \cite{Mermilliod:97}, and
we have used the color/temperature calibrations for giant stars from
\cite{Ramirez:05} for 13 different photometric indices. Interstellar
reddening has been ignored here in view of the close distance to the
star (13.8~pc), but we have accounted for the very small metallicity
correction in each calibration based on the discussion above. The
results are collected in Table~\ref{tab:photometry}, where the
uncertainty of each temperature estimate includes the contribution
from photometric errors as well as the statistical uncertainty of the
calibration, added in quadrature. The weighted average of these
determinations is $T_{\rm eff} = 4754 \pm 17$~K, although we prefer
100~K as a more realistic error to account for unquantified
systematics.  The spectroscopic and photometric temperature estimates
thus differ by only 100~K, and we adopt here the compromise value of
$T_{\rm eff} = 4800 \pm 100$~K.

Two additional properties of the star that can be determined very
accurately are the absolute visual magnitude and the linear
radius. The absolute magnitude follows from $V = 3.213 \pm 0.007$
\citep{Mermilliod:97} and our parallax for the system
(Table~\ref{tab:elements}), and is $M_V = 2.521 \pm 0.014$. The
angular diameter of \gamcep\ has been measured directly with high
precision by \cite{Nordgren:99} using the Navy Prototype Optical
Interferometer, and is $\phi = 3.24 \pm 0.03$~mas (limb-darkened
value). Combined once again with our parallax, this measurement yields
the linear radius as $R = 4.790 \pm 0.052$~R$_{\sun}$, which has a
formal precision just over 1\%.

\begin{figure}
\vskip -0.5in
\epsscale{1.00}
\plotone{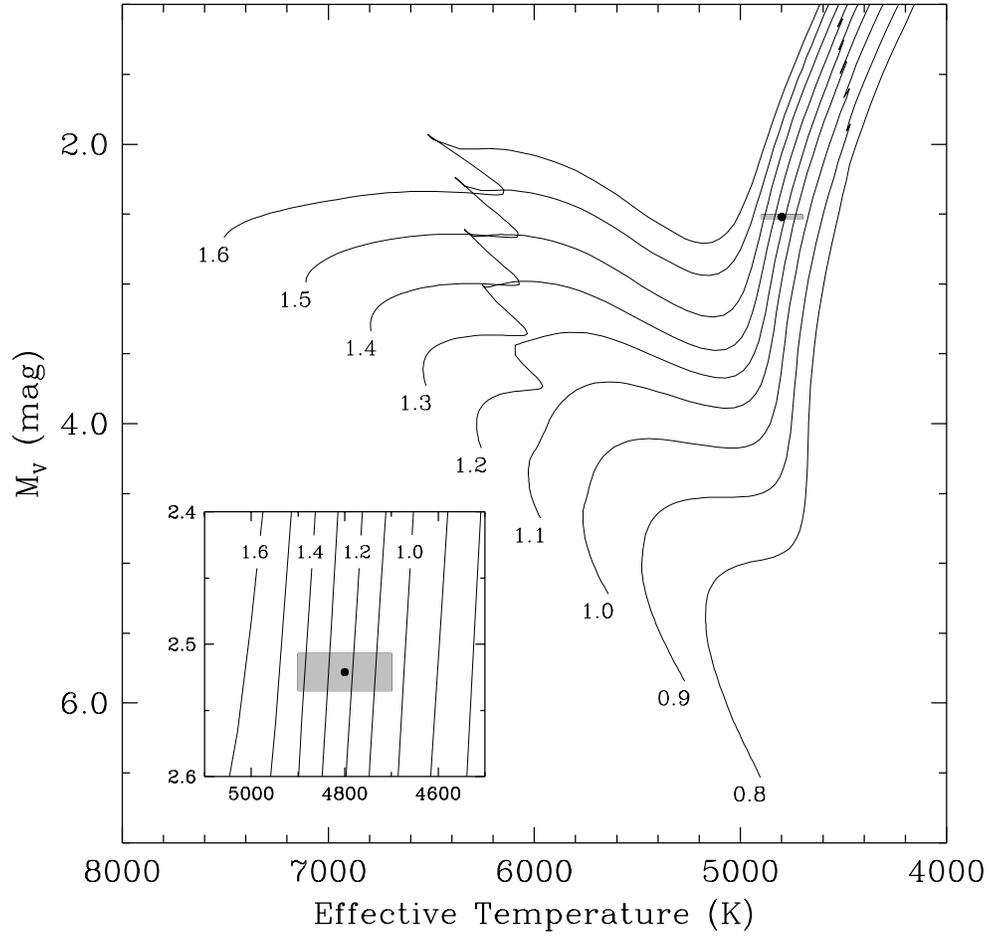}
\vskip -1.3in

 \figcaption[]{Evolutionary tracks from the calculations by
 \cite{Yi:01} and \cite{Demarque:04} in the absolute visual magnitude
 vs.\ effective temperature plane. The metallicity adopted is [Fe/H] =
 +0.01, the weighted average of all spectroscopic
 determinations. Masses are labeled in solar units, and the dot with
 the shaded error box represents the measurements for \gamcep. An
 enlargement is shown in the inset.\label{fig:tracks1}}

\end{figure}

\begin{figure}
\vskip -0.5in
\epsscale{1.00}
\plotone{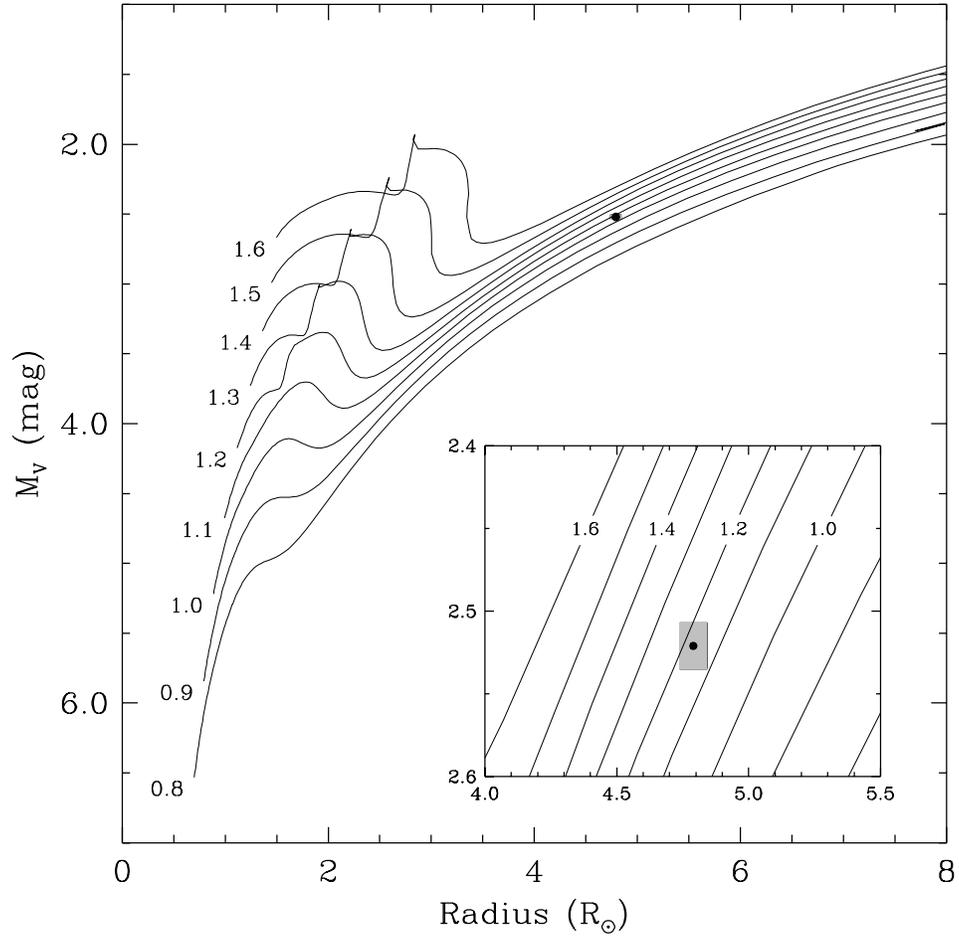}
\vskip -1.3in

 \figcaption[]{Same as Figure~\ref{fig:tracks1} but in the $M_V$ vs.\
 radius plane. The constraint on the mass of \gamcep\ is seen to be
 much tighter.\label{fig:tracks2}}

\end{figure}

In Figure~\ref{fig:tracks1} we compare the measured temperature and
absolute visual magnitude of the star with evolutionary tracks from
the series by \cite{Yi:01} and \cite{Demarque:04} for the composition
established above.  Tracks are labeled with the mass in solar units.
The star is seen to be in the first ascent of the giant branch. The
shaded error box is shown more clearly in the inset, which suggests a
mass for \gamcep\ slightly over 1.2~M$_{\sun}$ and an uncertainty in
that value determined almost entirely by the temperature error at this
fixed metallicity.  If the radius is used instead of the temperature,
the constraint on the mass is considerably improved because of the
smaller relative error of $R$. This is shown in
Figure~\ref{fig:tracks2}, which indicates a mass also close to
1.2~M$_{\sun}$, consistent with the previous figure. In both cases the
uncertainty in the metallicity is also important, as a change in
[Fe/H] shifts the tracks essentially horizontally.

\begin{figure}
\vskip -0.2in
\epsscale{1.00}
\plotone{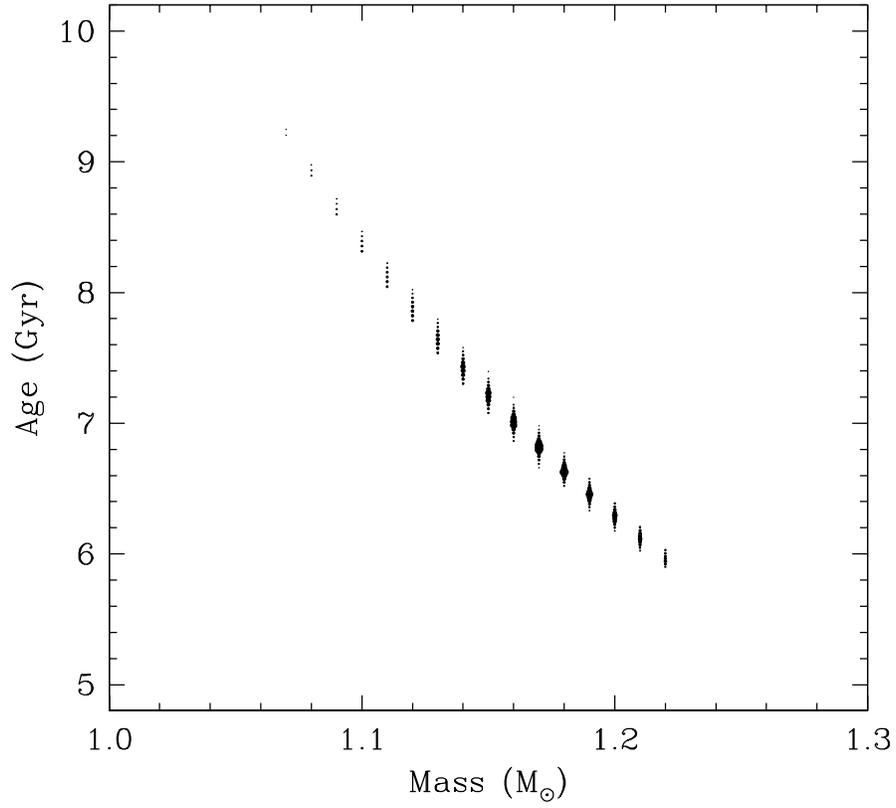}
\vskip -1.8in

 \figcaption[]{Theoretical mass and age combinations that are
 consistent with the four measured properties of \gamcep\ ($T_{\rm
 eff}$, [Fe/H], $M_V$, and $R$) within their errors. The best fit is
 for $M_{\rm Aa} = 1.18_{-0.11}^{+0.04}$~M$_{\sun}$ and an age of
 $6.6_{-0.7}^{+2.6}$~Gyr. The larger point sizes indicate a closer
 match.\label{fig:agemass}}

\end{figure}

The optimal value of the mass is one that yields the best simultaneous
match to the four measured quantities ($T_{\rm eff}$, [Fe/H], $M_V$,
and $R$) within their stated errors. To determine this value, as well
as its uncertainty, we computed by interpolation evolutionary tracks
in a fine grid for a range of masses and also a range of metallicities
within the observational uncertainty of [Fe/H]. At each point along
the tracks we compared the predicted stellar properties with the
measurements, and recorded all models that agree with the observations
within their errors. All such models are displayed in
Figure~\ref{fig:agemass} in a mass/age diagram. It is seen that at
each mass the range of allowed ages is very narrow. The best match is
for a mass of $M_{\rm Aa} = 1.18_{-0.11}^{+0.04}$~M$_{\sun}$, and the
corresponding evolutionary age is $6.6_{-0.7}^{+2.6}$~Gyr. All four
measured quantities are reproduced to well within their errors (better
than 0.3$\sigma$), an indication that they are mutually
consistent. The surface gravity predicted by the best model is $\log g
= 3.15$. An independent age estimate was obtained by \cite{Saffe:05}
based on the chromospheric activity indicator $\log R_{\rm
HK}^{\prime} = -5.32$. Their result (6.39~Gyr) using the calibration
by \cite{Rocha-Pinto:98} agrees very well with ours formally, although
chromospheric ages for older objects tend to be rather
uncertain\footnote{An additional age estimate by \cite{Saffe:05} based
on the calibration by \cite{Donahue:93} gave the value 14.78~Gyr for
\gamcep, which, however, is older than the age of the Universe.}.

There are significant differences between our mass and other recent
estimates. Almost all of them rely on evolutionary models and use
different combinations of observational constraints.  For example,
\cite{Fuhrmann:04} derived a value of 1.59~M$_{\sun}$ with a formal
error less than 10\%, from a fit to the effective temperature and
bolometric magnitude (derived using the {\it Hipparcos\/} parallax)
for a fixed metallicity that is higher than ours (see
Table~\ref{tab:specmet}). This mass estimate was adopted by
\cite{Hatzes:03} to infer the minimum mass of the substellar companion
to \gamcep.  \cite{Affer:05} obtained an even larger primary mass of
1.7~M$_{\sun}$ (no uncertainty given) from a fit to their own $T_{\rm
eff}$ and $M_V$ (also based on the {\it Hipparcos\/} parallax), using
their [Fe/H] determination that is again higher than
ours. \cite{AllendePrieto:99} used $M_V$ and $B-V$ directly and
inferred $M_{\rm Aa} = 1.33$~M$_{\sun}$, but apparently made no use of
any measured metallicity. A lower mass than ours ($M_{\rm Aa} = 1.0
\pm 0.2$~M$_{\sun}$) was derived by \cite{Luck:95} from the luminosity
and temperature they determined for \gamcep, along with their [Fe/H]
value, which is close to solar. \cite{Glebocki:72} obtained
1.5~M$_{\sun}$ employing a similar method, but adopted a metallicity
much lower than ours. Except for the latter study, the evolutionary
models used by most of these authors are similar enough that the
differences in mass must be due in large part to the observational
constraints, particularly the temperature and metallicity. We note
also that none of these studies have made use of the measured angular
diameter of the star, which appears to be very accurate.  An entirely
different approach was followed by \cite{Gratton:82}, who inferred a
mass from their spectroscopic $\log g$ determination along with a
linear radius derived from surface brightness relations ($R =
6.5$~R$_{\sun}$). Their value is $M_{\rm Aa} = 0.89 \pm
0.19$~M$_{\sun}$.

\begin{figure}
\vskip -0.2in
\epsscale{1.00}
\plotone{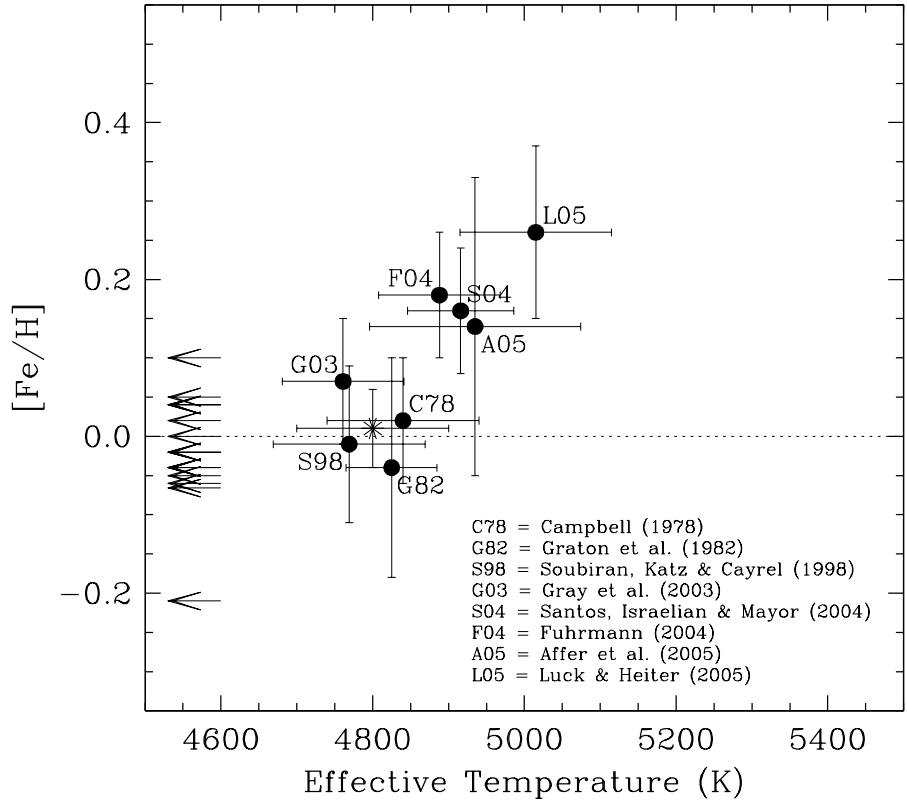}
\vskip -1.8in

 \figcaption[]{Effective temperature and metallicity determinations
 for \gamcep\ from only the studies that measured both quantities
 spectroscopically (filled circles). There is an apparent correlation
 between [Fe/H] and $T_{\rm eff}$. The values adopted in this paper
 are indicated with an asterisk. The arrows represent other
 spectroscopic abundance determinations that do not have a
 corresponding spectroscopic temperature
 measurement.\label{fig:teffmet}}

\end{figure}

The larger estimates of $M_{\rm Aa}$ tend to be those based on hotter
temperatures and also higher metallicities. Furthermore, a look at
Table~\ref{tab:specmet} shows that while most of the metallicity
determinations for \gamcep\ are close to solar, all the higher values
have been reported only in the last few years, and they tend to go
together with hotter temperatures (see Figure~\ref{fig:teffmet}). We
note in this connection that \gamcep\ has been considered a member of
a group of evolved stars displaying CN bands that are stronger than
usual \citep[`strong-CN stars', or `very-strong-lined stars'; see,
e.g.,][]{Spinrad:69, Keenan:87}. It was classified by the latter
authors as K1~III-IV~CN~1, which corresponds only to a marginal
strong-CN star. These objects have had a controversial history,
occasionally having been considered to be super-metal-rich. Other
studies have disputed this, however. For example, some of the giants
in the open cluster M67 have been found to have strong CN features
even though the chemical composition of this cluster is believed to be
essentially solar \citep[see, e.g.,][]{Luck:95}. We refer the reader
to the latter work (and references therein) for an excellent summary
of the subject and a list of possible explanations for the CN
phenomenon.

As tempting as it may be to place higher confidence in some of the
more recent [Fe/H] studies that have found a metal-rich composition
for \gamcep, it is difficult to ignore the large body of equally
careful determinations yielding a composition closer to solar. This
includes the recent work of \cite{Franchini:04}, which not only gives
a slightly subsolar metallicity but also happens to have the smallest
formal uncertainty; their result is [Fe/H] = $-0.066 \pm 0.034$. As a
test we repeated the comparison with stellar evolution models
described earlier, but adopting the spectroscopic temperature and
metallicity determinations of each of the recent studies that give
super-solar abundances. In no case did we find a model that is
simultaneously consistent with all four of the quantities $T_{\rm
eff}$, [Fe/H], $M_V$ and $R$, within their uncertainties. We are led
to conclude, therefore, that the chemical composition of \gamcep\ is
\emph{not} significantly higher than solar, and we adopt in the
following the mass we determined above with the most conservative of
the asymmetric error bars: $M_{\rm Aa} = 1.18 \pm 0.11$~M$_{\sun}$.

With this value and the mass function from our orbital solution the
mass of the unseen stellar companion is $M_{\rm B} = 0.362 \pm
0.022$~M$_{\sun}$, where the error is computed from the full
covariance matrix resulting from our fit (including cross-terms) and
accounts also for the primary mass uncertainty, which represents the
dominant contribution. Thus the secondary is most likely a late-type
star\footnote{For completeness we mention here two alternate
possibilities, although we consider them much less likely. One is that
the secondary is a white dwarf. In this case its low mass would make
it a helium-core white dwarf, which are the products of binary
evolution involving mass transfer through Roche-lobe overflow. Not
only is it difficult to see how the substellar companion could have
survived in this environment (unless it formed later, perhaps from
remnant material), but there also appears to be no evidence of a
(presumably hot) white dwarf in ultraviolet spectra of \gamcep. The
other possibility is that the companion is itself a closer binary
composed of smaller main-sequence stars. In this case their combined
brightness would be significantly less than that of a single M4 star
of the same mass, making it more difficult to detect \gamcep~B.} of
spectral type approximately M4. The angular semimajor axis of the
relative orbit between the primary and secondary becomes $a''_{\rm AB}
= 1.382 \pm 0.047$ arcsec, which corresponds to $19.02 \pm 0.64$ AU.

With the secondary mass known, it is of interest to compute its
brightness relative to the primary in order to assess the chance of
detecting it directly, most likely in the infrared.  The brightness
measurements of the primary itself in the near infrared are rather
uncertain because the star saturated the 2MASS detectors (see
Table~\ref{tab:photometry}). From our best model fits we derive
absolute magnitudes of $M_H \simeq 0.28$ and $M_K \simeq 0.19$ in the
Johnson system, which are actually consistent with the values inferred
from the 2MASS photometry within their large errors.  The brightness
of the secondary star may be estimated also from stellar evolution
models. For this we have used the calculations by \cite{Baraffe:98},
since those of \cite{Yi:01} are not intended for low-mass stars. For
the age we established above we obtain $M_V \simeq 10.92$, $M_H \simeq
6.83$, and $M_K \simeq 6.56$, which we have placed on the same
photometric system as \cite{Yi:01} following the prescription by
\cite{Bessell:88}. Thus, the secondary is expected to be $\sim$8.4 mag
fainter than \gamcep~A in $V$, $\sim$6.6 mag fainter in $H$, and
$\sim$6.4 mag fainter in $K$.\footnote{The brightness of the secondary
in $V$ may be overestimated by up to 0.5 mag due to the possibility of
missing opacities in the models \citep[see, e.g.,][]{Delfosse:00,
Chabrier:05}, which would affect the optical the most.}

The orbital elements in Table~\ref{tab:elements} allow the relative
position of the unseen secondary to be predicted. We note, however,
that the scale of the relative orbit still depends critically on the
assumed primary mass as $(M_{\rm Aa}+M_{\rm B})^{1/3}$, in which the
secondary mass itself scales as $(M_{\rm Aa}+M_{\rm B})^{2/3}$. As
seen earlier $M_{\rm Aa}$ is quite sensitive to the adopted
temperature and metallicity.  A dynamical (hypothesis-free) estimate
of the masses of both stars and a direct measure of the semimajor axis
$a''_{\rm AB}$ will be possible once \gamcep~B is detected and its
path around the primary measured over at least a portion of the
orbital cycle.

\section{The mass of the planetary companion}
\label{sec:planet}

The reflex motion of the primary star along the line of sight in
response to the putative substellar companion leads to a mass function
of $f(M_p) = (1.83 \pm 0.32) \times 10^{-9}$~M$_{\sun}$ from our
orbital fit. With the adopted value of $M_{\rm Aa}$ this corresponds
to $M_p \sin i_{\rm A} = 1.43 \pm 0.13$~M$_{\rm Jup}$, which is only
slightly smaller than the value $M_p \sin i_{\rm A} = 1.7 \pm
0.4$~M$_{\rm Jup}$ reported by \cite{Hatzes:03}. The difference is due
almost entirely to the choice of primary mass, for which they used
$M_{\rm Aa} = 1.59$~M$_{\sun}$.

The perturbation on the primary star on the plane of the sky caused by
the substellar companion is expected to be small, although it depends
obviously on the unknown inclination angle $i_{\rm A}$ (and through it
on $M_p$). Given that the {\it Hipparcos\/} measurements are fairly
precise, we attempted to determine this astrometric wobble
simultaneously with the other elements by incorporating additional
adjustable parameters into the model. Four of the elements of this
astrometric orbit are already known from spectroscopy ($P_{\rm A}$,
$e_{\rm A}$, $\omega_{\rm Aa}$, and $T_{\rm A}$). The remaining three
are the angular scale (semimajor axis) of the orbit of the primary
around its center of mass with the planet ($a''_{\rm Aa}$), the
inclination angle of the planetary orbit ($i_{\rm A}$), and the
position angle of the ascending node ($\Omega_{\rm A}$, J2000). Since
spectroscopy gives the projected linear semimajor axis ($a_{\rm Aa}
\sin i_{\rm A}$), and the parallax is a known function of other
elements (see eq.~\ref{eq:eq1}), we take advantage of the redundancy
to eliminate the angular semimajor axis $a''_{\rm Aa}$ as an
adjustable parameter, given that it can be expressed as
 \begin{equation}
a''_{\rm Aa} = a''_{\rm A} {P_{\rm A} \over P_{\rm AB}} {K_{\rm Aa}
\over K_{\rm A}} \sqrt{{1 - e_{\rm A}^2 \over 1 - e_{\rm AB}}} {\sin
i_{\rm AB} \over \sin i_{\rm A}}~.
\label{eq:eq2}
 \end{equation}
A solution with a total of 25 adjustable parameters did not yield a
statistically significant detection of the astrometric wobble: the
best fit corresponded to an inclination angle 19\arcdeg\ from face-on,
implying a semimajor axis $a''_{\rm Aa} = 0.46 \pm 0.36$~mas and a
planet mass around 4~M$_{\rm Jup}$. 

\begin{figure}
\vskip -0.5in
\epsscale{1.00}
\plotone{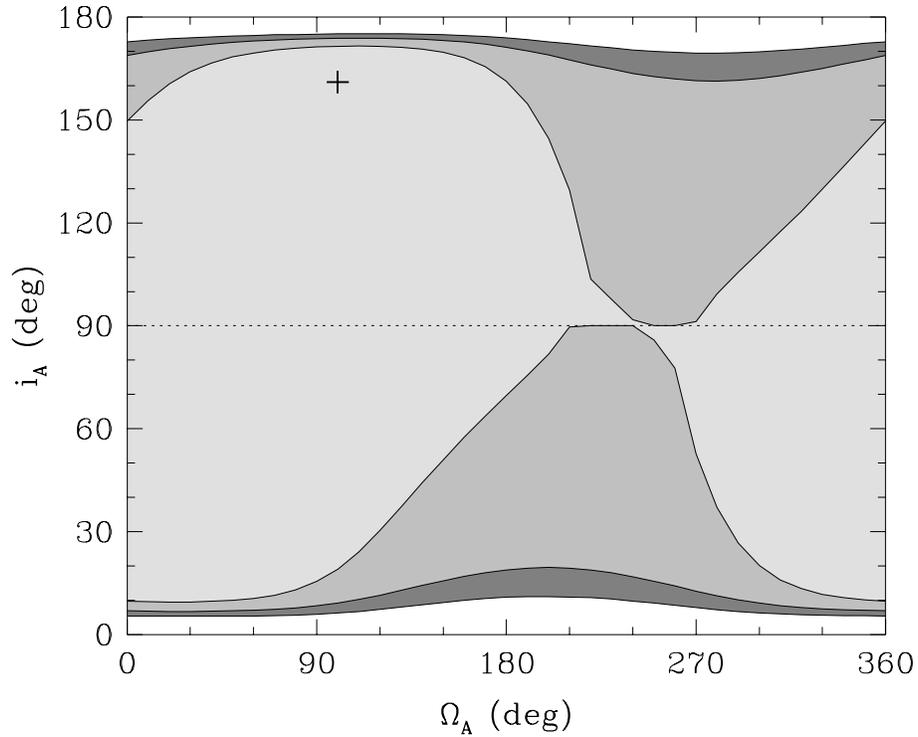}
\vskip -1.6in

 \figcaption[]{Confidence levels of orbital fits in the $i_{\rm
 A}$--$\Omega_{\rm A}$ space of parameters, describing the wobble of
 \gamcep~A on the plane of the sky in response to the pull from the
 substellar companion. We use these fits to place an upper limit on
 the mass $M_p$ of the substellar companion. The light gray area
 represents solutions for fixed values of these two parameters that
 can be ruled out only at the $\sim$68\% confidence level (1$\sigma$)
 or less. The plus sign corresponds to the best fit, which however
 does not give a statistically significant result (see text). The
 middle shade of gray corresponds to fits ruled out at confidence
 levels between 1$\sigma$ and 2$\sigma$, and yields an upper limit on
 $M_p$ of 13.3~M$_{\rm Jup}$. The dark gray area corresponds to fits
 ruled out at confidence levels between 2$\sigma$ and 3$\sigma$. The
 outer edge of this region yields a minimum inclination angle of
 $4\fdg9$, and an upper limit on $M_p$ of 16.9~M$_{\rm
 Jup}$. \label{fig:planetmass}}

\end{figure}

In order to place a meaningful upper limit on $M_p$ we explored the
full range of possible values of $i_{\rm A}$ and $\Omega_{\rm A}$ to
identify the area of parameter space where the solutions become
inconsistent with the observational errors. For each pair of fixed
values of $i_{\rm A}$ and $\Omega_{\rm A}$ we solved for the other 23
parameters of the fit as usual. A false alarm probability can be
attached to the $\Delta\chi^2$ (increase in $\chi^2$ compared to the
minimum) associated with each of these solutions.  In this way we may
determine the minimum value of $\sin i_{\rm A}$ (highest value of
$M_p$) for a given confidence level. This is illustrated in
Figure~\ref{fig:planetmass}, where we show the region of parameter
space in the two variables of interest along with confidence
contours. The light gray area corresponds to solutions that can only
be ruled out at confidence levels up to 1$\sigma$ ($\sim$68\%), and
includes our best fit mentioned above (indicated with a plus
sign). The middle gray area is the region between 1$\sigma$ and
2$\sigma$, and the dark gray area corresponds to confidence levels
between 2$\sigma$ and 3$\sigma$. At the 2$\sigma$ level ($\sim$95\%
confidence) the observations rule out companion masses larger than
13.3~M$_{\rm Jup}$ (or inclination angles less than $6\fdg2$ from
face-on), which would induce reflex motions on the primary with a
semiamplitude of at least 1.5~mas. This mass corresponds roughly to
the conventional boundary between planetary and brown-dwarf masses.
At a higher confidence level of 3$\sigma$ (99.73\%) the mass limit is
16.9~M$_{\rm Jup}$ (or $i_{\rm A} > 4\fdg9$), which would produce a
wobble with a semiamplitude of about 1.8~mas. There is little doubt,
therefore, that the companion is substellar.

\section{Discussion and concluding remarks}
\label{sec:discussion}

In their paper \cite{Hatzes:03} attempted to place limits on the mass
of the substellar companion in a different way by using their measured
projected rotational velocity for \gamcep\ ($v \sin i = 1.5 \pm
1.0$~\kms) along with the period they determined for the variation of
the \ion{Ca}{2}~$\lambda$8662 emission-line index ($781 \pm 116$~days)
and the estimated radius of the star \citep[$R=4.66$~R$_{\sun}$,
adopted from][]{Fuhrmann:04}. They relied on two assumptions: that the
spin axis of the star is parallel to the axis of the planetary orbit,
and that the period of variation of the \ion{Ca}{2} index represents
the true rotation period of the star. The comparison between the
measured $v \sin i$ and the expected equatorial rotational velocity
($v_{\rm eq}$) then gives limits on $\sin i$ (or $\sin i_{\rm A}$ in
our notation).  As it turns out, however, there is a mathematical
error in their calculation of $v_{\rm eq}$: they reported $v_{\rm eq}
= 4.9$~\kms, while the correct value is 0.3~\kms\ \citep[see
also][]{Walker:92}. Since this is \emph{smaller} than their $v \sin
i$, no limits can be placed on $\sin i_{\rm A}$ in this way. Their
statement on the probable mass range of the planetary companion is
therefore not valid.  Either the measured $v \sin i$ is overestimated,
or the period of the \ion{Ca}{2} variations is not the true rotation
period of the star. The former explanation is perhaps supported by a
measurement by \cite{Gray:85}, who gave $v \sin i = 0.0 \pm 0.8$~\kms\
(along with a sizeable radial-tangential macroturbulence of
$\zeta_{\rm RT} = 4.2 \pm 0.6$~\kms, which can effect rotational
velocity measurements if not properly accounted for). The study by
\cite{deMedeiros:99} reported $v \sin i < 1.0$~\kms.  Alternatively,
the rotation period would have to be considerably shorter than 781
days ($\approx$100--500 days), and another explanation would have to
be found for the variations in the emission-line index.  Our dynamical
constraint on $M_p$ thus shows for the first time that the companion
is substellar in nature, although a mass in the brown dwarf regime (as
opposed to the planetary regime) cannot be completely ruled out with
the present observations.

\gamcep\ is one of more than two dozen examples of substellar
companions found in stellar binaries \citep[see,
e.g.,][]{Raghavan:06}. Such systems have attracted considerable
interest in recent years, and numerical studies have been carried out
specifically for the case of \gamcep\ to assess not only the dynamical
stability of the orbit of the substellar companion
\citep[e.g.,][]{Dvorak:03, Solovaya:04, Haghighipour:06}, but also the
stability of the orbits of other (possibly Earth-like) planets that
might be present in the habitable zone of the primary
star. \cite{Thebault:04} have also investigated the conditions under
which the substellar companion may form through core accretion in the
binary environment.  With a relative semimajor axis for the planet
orbit of $a_{{\rm Aa}-p} = 1.94 \pm 0.06$~AU (for an adopted primary
mass $M_{\rm Aa} = 1.18 \pm 0.11$~M$_{\sun}$), the size of that orbit
is only 9.8 times smaller than the size of the binary orbit (19.02~AU;
see Table~\ref{tab:elements}), currently the lowest value among the
known exoplanets in binaries\footnote{The slightly smaller orbit size
ratio compared to the value of $\sim$11 given by \cite{Raghavan:06} is
largely due to the significant improvement in the elements of the
binary orbit in the present work.}. Orbit stability depends quite
strongly on the parameters of the binary system, in particular the
semimajor axis and eccentricity, as well as on the masses of the
components. The dynamical studies mentioned above have all had to make
do with the rather poorly determined binary properties and also often
inconsistent results from various authors.

\cite{Holman:99} have derived a simple empirical formula for computing
the maximum value of the semimajor axis of a stable planetary orbit
(``critical'' semimajor axis, $a_{\rm crit}$) in a coplanar S-type
planet-binary system.  \cite{Haghighipour:06} pointed out in his study
that the uncertainty in the binary orbital elements made for a very
large parameter space to be explored numerically for
\gamcep. Furthermore, the inclination of the binary orbit was unknown
at the time and therefore so was the mass of the secondary star. As a
result, he was only able to provide a rather wide range of critical
semimajor axes as a function of the adopted binary eccentricity (see
his Figure~1). With the present study that situation has changed, and
the critical semimajor axis can now be computed directly with a
relatively small formal uncertainty. We obtain $a_{\rm crit} = 3.61
\pm 0.36$~AU, which is considerably larger than the semimajor axis of
the planet orbit, implying the latter is stable if coplanar with the
binary.

The combination of classical as well as high-precision radial velocity
measurements of \gamcep\ with ground- and space-based astrometry have
allowed a significant improvement in the binary orbital elements (and
a first determination of the inclination angle) as well as a better
knowledge of the stellar masses. Nevertheless, the secondary star
remains unseen. Even though the predicted angular separation of
\gamcep~B (0\farcs84 for 2007.0; 0\farcs99 for 2009.0) is not
particularly challenging, the 8-magnitude brightness difference in the
visual band relative to the glaringly bright primary explains all
negative results (e.g., the speckle interferometry attempts by Mason
et al.\ 2001, as well as the imaging by Hatzes et al.\ 2003). We
expect the contrast to be much more favorable in the near infrared
($\Delta m \sim 6.4$ in $K$), and that this detection should not be
very difficult at those wavelengths with adaptive optics on a large
telescope. Such measurements of the relative position would allow a
dynamical determination of the mass of both stars, free from
assumptions.

\acknowledgements

Thanks are due to D.\ W.\ Latham for obtaining the spectroscopic
observations of \gamcep\ used here, and for helpful discussions on
zero-point corrections. We are also grateful to R.\ Neuh\"auser for
bringing this star to our attention, and to S.\ Urban and G.\
Gontcharov for providing ground-based catalog positions for
\gamcep. An anonymous referee is thanked for helpful comments.
Partial support for this work from NSF grant AST-0406183, NASA's
MASSIF SIM Key Project (BLF57-04), and NASA Origins grant NNG04LG89G
is acknowledged.  This research has made use of the SIMBAD database,
operated at CDS, Strasbourg, France, of NASA's Astrophysics Data
System Abstract Service, and of data products from the Two Micron All
Sky Survey, which is a joint project of the University of
Massachusetts and the Infrared Processing and Analysis
Center/California Institute of Technology, funded by NASA and the NSF.


\clearpage

\begin{deluxetable}{lcccccc}
\tabletypesize{\scriptsize}
\tablewidth{0pt}

\tablecaption{Heliocentric radial velocity measurements for \gamcep\
derived in this work as well as others collected from the literature,
all placed on the CfA reference frame.\label{tab:rvcfa}}

\tablehead{\colhead{HJD} & \colhead{} & \colhead{Orbital} &
           \colhead{RV\tablenotemark{a}} & 
           \colhead{$\sigma_{\rm RV}$\tablenotemark{b}} & 
           \colhead{O$-$C} & \colhead{} \\
           \colhead{~~~(2,400,000+)~~~} & \colhead{Year} &
           \colhead{Phase} & \colhead{($\kms$)} &
           \colhead{($\kms$)} & \colhead{($\kms$)} & \colhead{Source}}
\startdata
 16039.739 \dotfill &  1902.7919  &  0.6701  &  $-$42.23  &  0.75  &  $+$0.68  &  1 \\
 16208.653 \dotfill &  1903.2544  &  0.6770  &  $-$42.43  &  0.75  &  $+$0.52  &  1 \\
 16241.612 \dotfill &  1903.3446  &  0.6784  &  $-$42.93  &  0.75  &  $+$0.03  &  1 \\
 17131.88  \dotfill &  1905.7820  &  0.7149  &  $-$41.57  &  0.56  &  $+$1.46  &  2 \\
 17146.88  \dotfill &  1905.8231  &  0.7155  &  $-$43.17  &  0.56  &  $-$0.13  &  2 \\
 17152.83  \dotfill &  1905.8394  &  0.7157  &  $-$43.97  &  0.56  &  $-$0.93  &  2 \\
 17178.247 \dotfill &  1905.9090  &  0.7168  &  $-$42.64  &  0.75  &  $+$0.40  &  3 \\
 17180.223 \dotfill &  1905.9144  &  0.7168  &  $-$42.99  &  0.75  &  $+$0.06  &  3 \\
 17467.480 \dotfill &  1906.7008  &  0.7286  &  $-$43.94  &  0.75  &  $-$0.86  &  3 \\
 17494.387 \dotfill &  1906.7745  &  0.7297  &  $-$42.25  &  0.75  &  $+$0.83  &  3 \\
 17853.407 \dotfill &  1907.7574  &  0.7444  &  $-$43.37  &  0.75  &  $-$0.28  &  3 \\
 23021.616 \dotfill &  1921.9072  &  0.9563  &  $-$47.58  &  1.88  &  $-$1.74  &  4 \\
 35109.0   \dotfill &  1955.0007  &  0.4519  &  $-$42.30  &  1.13  &  $+$0.81  &  5 \\
 41496.971 \dotfill &  1972.4900  &  0.7138  &  $-$41.34  &  1.13  &  $+$1.69  &  6 \\
 41497.972 \dotfill &  1972.4927  &  0.7138  &  $-$42.18  &  1.13  &  $+$0.85  &  6 \\
 41642.597 \dotfill &  1972.8887  &  0.7197  &  $-$43.78  &  1.13  &  $-$0.72  &  6 \\
 41642.602 \dotfill &  1972.8887  &  0.7197  &  $-$43.70  &  1.13  &  $-$0.64  &  6 \\
 41642.606 \dotfill &  1972.8887  &  0.7197  &  $-$44.13  &  1.13  &  $-$1.07  &  6 \\
 41643.590 \dotfill &  1972.8914  &  0.7198  &  $-$45.12  &  1.13  &  $-$2.06  &  6 \\
 41643.618 \dotfill &  1972.8915  &  0.7198  &  $-$44.36  &  1.13  &  $-$1.30  &  6 \\
 41644.720 \dotfill &  1972.8945  &  0.7198  &  $-$44.16  &  1.13  &  $-$1.10  &  6 \\
 42203.973 \dotfill &  1974.4257  &  0.7427  &  $-$43.19  &  1.13  &  $-$0.11  &  6 \\
 42204.995 \dotfill &  1974.4285  &  0.7428  &  $-$43.80  &  1.13  &  $-$0.72  &  6 \\
 42334.781 \dotfill &  1974.7838  &  0.7481  &  $-$42.24  &  1.13  &  $+$0.88  &  6 \\
 42335.842 \dotfill &  1974.7867  &  0.7481  &  $-$43.18  &  1.13  &  $-$0.06  &  6 \\
 42336.762 \dotfill &  1974.7892  &  0.7482  &  $-$42.58  &  1.13  &  $+$0.54  &  6 \\
 43396.0   \dotfill &  1977.6893  &  0.7916  &  $-$42.59  &  0.47  &  $+$0.78  &  7 \\
 52099.9707\dotfill &  2001.5194  &  0.1485  &  $-$45.08  &  0.23  &  $+$0.00  &  8 \\
 53275.7782\dotfill &  2004.7386  &  0.1967  &  $-$44.72  &  0.23  &  $-$0.17  &  8 \\
 53337.6691\dotfill &  2004.9081  &  0.1992  &  $-$44.60  &  0.23  &  $-$0.07  &  8 \\
\enddata
\tablenotetext{a}{~Includes offsets as listed in Table~\ref{tab:offsets}.}
\tablenotetext{b}{~Includes scale factors described in the text.}
\tablecomments{Sources are: 1. \cite{Frost:03}; 2. \cite{Slipher:05};
3. \cite{Kustner:08}; 4. \cite{Harper:34}; 5. \cite{Boulon:57}; 6.
\cite{Snowden:05}; 7. \cite{Kjaergaard:81}; 8. This paper.}

\end{deluxetable}

\clearpage

\begin{deluxetable}{ccccc|ccccc}
\tabletypesize{\scriptsize}
\tablewidth{0pt}

\tablecaption{Ground-based positional measurements of \gamcep\ from
transit circle and photographic programs, on the International
Celestial Reference Frame.\label{tab:astrometry}}

\tablehead{\colhead{R.A.} & \colhead{Orbital} & \colhead{R.A.\ (J2000)} &
           \colhead{$\sigma_{\alpha}$\tablenotemark{a}} & \colhead{O$-$C} &
           \colhead{Dec.} & \colhead{Orbital} & \colhead{Dec.\ (J2000)} &
           \colhead{$\sigma_{\delta}$\tablenotemark{a}} & \colhead{O$-$C} \\
           \colhead{Epoch} & \colhead{Phase} & \colhead{(hh:mm:ss.ssss)} &
           \colhead{(mas)} & \colhead{(mas)} &
           \colhead{Epoch} & \colhead{Phase} & \colhead{(dd:mm:ss.sss)} &
           \colhead{(mas)} & \colhead{(mas)}}
\startdata
  1898.06 & 0.5992 & 23:39:22.9400 &  463    &  $+$122      &  1898.06 & 0.5992 &  $+$77:37:41.850  &  587    &  $+$120    \\
  1900.20 & 0.6313 & 23:39:22.8942 &  256    &  \phn$+$85   &  1901.19 & 0.6461 &  $+$77:37:42.320  &  171    &  $+$157    \\
  1905.37 & 0.7087 & 23:39:22.8527 &  253    &  $+$287      &  1905.03 & 0.7036 &  $+$77:37:42.953  &  262    &  $+$217    \\
  1907.87 & 0.7461 & 23:39:22.6199 &  225    &  $-$154      &  1907.98 & 0.7478 &  $+$77:37:42.974  &  183    &  $-$179    \\
  1907.88 & 0.7463 & 23:39:22.5343 &  225    &  $-$426      &  1907.88 & 0.7463 &  $+$77:37:42.963  &  188    &  $-$221    \\
  1911.70 & 0.8035 & 23:39:22.5729 &  476    &  \phn$-$72   &  1911.70 & 0.8035 &  $+$77:37:43.453  &  420    &  $-$261    \\
  1918.70 & 0.9083 & 23:39:22.5466 &  368    &  $+$355      &  1918.70 & 0.9083 &  $+$77:37:44.393  &  285    &  $-$137    \\
  1929.89 & 0.0759 & 23:39:22.2488 &  332    &  \phn$+$43   &  1929.89 & 0.0759 &  $+$77:37:46.203  &  295    &  $+$262    \\
  1940.91 & 0.2409 & 23:39:22.0967 &  138    &  $+$121      &  1940.91 & 0.2409 &  $+$77:37:47.945  &  154    &  $+$100    \\
  1945.48 & 0.3093 & 23:39:21.9705 &  123    &  $-$136      &  1945.48 & 0.3093 &  $+$77:37:48.832  &  164    &  $+$194    \\
  1952.70 & 0.4174 & 23:39:21.8727 &  \phn94 &  \phn$+$90   &  1952.70 & 0.4174 &  $+$77:37:49.622  &  115    &  $-$285    \\
  1957.67 & 0.4918 & 23:39:21.6400 &  330    &  $-$321      &  1957.67 & 0.4918 &  $+$77:37:50.911  &  285    &  $+$205    \\
  1979.28 & 0.8154 & 23:39:21.2509 &  110    &  \phn$+$42   &  1979.36 & 0.8166 &  $+$77:37:53.547  &  124    &  $-$124    \\
  1984.71 & 0.8967 & 23:39:21.1435 &  100    &  $+$127      &  1984.71 & 0.8967 &  $+$77:37:54.386  &  130    &  \phn$-$15 \\
  1985.23 & 0.9045 & 23:39:21.1245 &  100    &  \phn$+$89   &  1985.22 & 0.9044 &  $+$77:37:54.332  &  130    &  \phn$+$17 \\
  1985.68 & 0.9113 & 23:39:21.0996 &  100    &  \phn$+$40   &  1985.68 & 0.9113 &  $+$77:37:54.469  &  130    &  \phn$-$39 \\
  1986.30 & 0.9206 & 23:39:21.1259 &  100    &  $+$139      &  1986.31 & 0.9207 &  $+$77:37:54.487  &  130    &  \phn$+$33 \\
  1986.76 & 0.9274 & 23:39:21.0801 &  100    &  \phn$+$82   &  1986.76 & 0.9274 &  $+$77:37:54.667  &  130    &  \phn$+$44 \\
  1987.51 & 0.9387 & 23:39:21.0507 &  100    &  \phn$-$50   &  1987.54 & 0.9391 &  $+$77:37:54.743  &  130    &  \phn$+$60 \\
  1987.71 & 0.9417 & 23:39:21.0636 &  100    &  \phn$+$68   &  1987.71 & 0.9417 &  $+$77:37:54.678  &  130    &  \phn$-$55 \\
  1988.36 & 0.9514 & 23:39:21.1137 &  100    &  $+$212      &  1988.36 & 0.9514 &  $+$77:37:54.728  &  109    &  \phn$+$28 \\
  1989.08 & 0.9622 & 23:39:21.0301 &  \phn83 &  \phn$+$87   &  1989.09 & 0.9623 &  $+$77:37:54.901  &  104    &  $+$140    \\
  1989.25 & 0.9647 & 23:39:21.0495 &  \phn83 &  \phn$+$93   &  1989.25 & 0.9647 &  $+$77:37:54.620  &  104    &  $-$150    \\
  1990.24 & 0.9796 & 23:39:21.0441 &  \phn83 &  $+$133      &  1990.24 & 0.9796 &  $+$77:37:54.826  &  104    &  \phn$-$59 \\
  1990.70 & 0.9864 & 23:39:20.9988 &  \phn83 &  \phn$+$24   &  1990.70 & 0.9864 &  $+$77:37:55.128  &  104    &  \phn$+$47 \\
  1990.75 & 0.9872 & 23:39:20.9900 &  \phn57 &  \phn$+$18   &  1990.75 & 0.9872 &  $+$77:37:55.090  &  \phn61 &  \phn\phn$+$7  \\
  1991.66 & 0.0008 & 23:39:20.9811 &  \phn66 &  \phn\phn$-$2&  1991.69 & 0.0013 &  $+$77:37:55.034  &  \phn96 &  $-$172    \\
  1991.87 & 0.0040 & 23:39:20.9973 &  \phn66 &  $+$136      &  1991.87 & 0.0040 &  $+$77:37:55.200  &  \phn96 &  \phn\phn$+$9  \\
  1993.27 & 0.0249 & 23:39:20.9700 &  \phn66 &  \phn$+$28   &  1993.26 & 0.0248 &  $+$77:37:55.470  &  \phn96 &  $+$187    \\
  1993.58 & 0.0296 & 23:39:20.9423 &  \phn66 &  \phn$-$68   &  1993.58 & 0.0296 &  $+$77:37:55.531  &  \phn96 &  \phn$+$83 \\
  1994.55 & 0.0441 & 23:39:20.9401 &  \phn66 &  \phn$-$41   &  1994.54 & 0.0439 &  $+$77:37:55.448  &  \phn96 &  $-$129    \\
  1994.75 & 0.0471 & 23:39:20.9412 &  \phn66 &  \phn$+$44   &  1994.75 & 0.0471 &  $+$77:37:55.672  &  \phn96 &  \phn$+$36 \\
\enddata
\tablenotetext{a}{Includes scale factors described in the text.}
\end{deluxetable}

\clearpage

\begin{deluxetable}{lc}
\tablewidth{0pt}

\tablecaption{Global orbital solution for \gamcep.\label{tab:elements}}

\tablehead{\colhead{~~~~~~~~~~~~~~~~~~Parameter~~~~~~~~~~~~~~~~~~} & \colhead{Value}}
\startdata
\sidehead{Adjusted quantities from outer orbit (A+B)} \\
\noalign{\vskip -9pt}
~~~~$P_{\rm AB}$ (days)\dotfill             &  24392~$\pm$~522\phn\phn \\
~~~~$P_{\rm AB}$ (yr)\dotfill               &  66.8~$\pm$~1.4\phn \\
~~~~$\gamma$ (\kms)\dotfill                 &  $-$42.958~$\pm$~0.047\phn\phs \\
~~~~$K_{\rm A}$ (\kms)\dotfill              &  1.925~$\pm$~0.014 \\
~~~~$e_{\rm AB}$\dotfill                    &  0.4085~$\pm$~0.0065 \\
~~~~$\omega_{\rm A}$ (deg)\dotfill          &  160.96~$\pm$~0.40\phn\phn \\
~~~~$T_{\rm AB}$ (HJD$-$2,400,000)\dotfill  &  48479~$\pm$~12\phn\phn\phn \\
~~~~$T_{\rm AB}$ (yr)\dotfill               &  1991.606~$\pm$~0.032\phn\phn\phn \\
~~~~$a''_{\rm A}$ (mas)\dotfill             &  324.6~$\pm$~8.4\phn\phn \\
~~~~$i_{\rm AB}$ (deg)\dotfill              &  118.1~$\pm$~1.2\phn\phn \\
~~~~$\Omega_{\rm AB}$ (deg)\dotfill         &  13.0~$\pm$~2.4\phn \\
\sidehead{Adjusted quantities from inner orbit (Aa+Ab)} \\
\noalign{\vskip -9pt}
~~~~$P_{\rm A}$ (days)\dotfill              &  902.8~$\pm$~3.5\phn\phn \\
~~~~$P_{\rm A}$ (yr)\dotfill                &  2.4717~$\pm$~0.0096 \\
~~~~$K_{\rm Aa}$ (\ms)\dotfill              &  27.1~$\pm$~1.5\phn \\
~~~~$e_{\rm A}$\dotfill                     &  0.113~$\pm$~0.058 \\
~~~~$\omega_{\rm Aa}$ (deg)\dotfill         &  63~$\pm$~27 \\
~~~~$T_{\rm A}$ (HJD$-$2,400,000)\dotfill   &  53146~$\pm$~72\phn\phn\phn \\
~~~~$T_{\rm A}$ (yr)\dotfill                &  2004.38~$\pm$~0.20\phn\phn\phn \\
\sidehead{Other adjusted quantities} \\
\noalign{\vskip -9pt}
~~~~$\Delta RV_1$ (\kms) [McDonald~I]\tablenotemark{a}\dotfill    &  $-$45.228~$\pm$~0.035\phn\phs \\
~~~~$\Delta RV_2$ (\kms) [McDonald~II]\tablenotemark{a}\dotfill   &  $-$45.424~$\pm$~0.035\phn\phs \\
~~~~$\Delta RV_3$ (\kms) [McDonald~III]\tablenotemark{a}\dotfill  &  $-$44.053~$\pm$~0.035\phn\phs \\
~~~~$\Delta RV_4$ (\kms) [CFHT]\tablenotemark{a}\dotfill          &  $-$44.483~$\pm$~0.035\phn\phs \\
~~~~$\Delta RV_5$ (\kms) [CfA]\tablenotemark{a}\dotfill           &  +1.13~$\pm$~0.12\phs \\
~~~~$\Delta\alpha^*$ (mas)\dotfill               &  +73.6~$\pm$~7.5\phn\phs \\
~~~~$\Delta\delta$ (mas)\dotfill                 &  +160.1~$\pm$~3.9\phn\phn\phs \\
~~~~$\Delta\mu_{\alpha}^*$ (mas~yr$^{-1}$)\dotfill   &  $-$16.0~$\pm$~1.1\phn\phs \\
~~~~$\Delta\mu_{\delta}$ (mas~yr$^{-1}$)\dotfill     &  +21.91~$\pm$~0.81\phn\phs \\
\sidehead{Derived quantities} \\
\noalign{\vskip -9pt}
~~~~R.A.\ (sec)\tablenotemark{b}~~~23$^{\rm h}$39$^{\rm m}$~\dotfill   & 21.0050~$\pm$~0.0023\phn \\
~~~~Dec.\ (arcsec)\tablenotemark{b}~~~+77\arcdeg 37\arcmin~\dotfill  &  55.241~$\pm$~0.004\phn \\
~~~~$\mu_{\alpha}^*$ (mas~yr$^{-1}$)\dotfill         &  $-$64.8~$\pm$~1.1\phn\phs \\
~~~~$\mu_{\delta}$ (mas~yr$^{-1}$)\dotfill           &  +149.09~$\pm$~0.81\phn\phn\phs \\
~~~~$\pi$ (mas)\dotfill                       &  72.70~$\pm$~0.39\phn \\
~~~~$a''_{\rm AB}$ (arcsec)\dotfill           &  1.382~$\pm$~0.047 \\
~~~~$a_{\rm AB}$ (AU)\dotfill                 &  19.02~$\pm$~0.64\phn \\
~~~~$f(M_{\rm B})$ (M$_{\sun}$)\dotfill       &  0.01371~$\pm$~0.00049 \\
~~~~$M_{\rm B}$ (M$_{\sun}$)\tablenotemark{c}\dotfill          &  0.362~$\pm$~0.022 \\
~~~~$f(M_p)$ ($10^{-9}$~M$_{\sun}$)\dotfill   &  1.83~$\pm$~0.32 \\
~~~~$M_p \sin i_{\rm A}$ (M$_{\rm Jup}$)\tablenotemark{c}\dotfill     &  1.43~$\pm$~0.13 \\
~~~~$a_{{\rm Aa}-p}$ (AU)\tablenotemark{c,d}\dotfill             &  1.94~$\pm$~0.06 \\
\enddata
\tablenotetext{a}{Offsets to be \emph{added} to the corresponding data sets in order
to place them on the Griffin system.}
\tablenotetext{b}{Coordinates of the barycenter (ICRF, J2000, epoch 1991.25).}
\tablenotetext{c}{Assumes a primary mass of $M_{\rm Aa} = 1.18 \pm
0.11$ M$_{\sun}$ (see \S\ref{sec:properties}).}
\tablenotetext{d}{Relative semimajor axis of the orbit of the substellar companion.}

\end{deluxetable}

\clearpage

\begin{deluxetable}{lccc}
\tabletypesize{\scriptsize}
\tablewidth{0pc}
\tablecaption{Spectroscopic determinations of the effective temperature
and metallicity for \gamcep\ from the literature.\label{tab:specmet}}
\tablehead{\colhead{} & \colhead{$T_{\rm eff}$\tablenotemark{a}} &
\colhead{[Fe/H]} & \colhead{$\log g$} \\
\colhead{~~~~~~~~~~~~~~~~~~~~~~~~~~~~Source~~~~~~~~~~~~~~~~~~~~~~~~~~~~} & \colhead{(K)} & \colhead{(dex)} &
\colhead{(dex)} }
\startdata

\cite{Herbig:66}\dotfill     &   4383\tablenotemark{b}  &   \phm{\tablenotemark{c}}+0.05\tablenotemark{c}\phs   &  \nodata          \\
\cite{Spite:66}\dotfill      &   \nodata                &   +0.02\phs          &  \nodata          \\
\cite{Spinrad:69}\dotfill    &   \nodata                &   \phm{:}+0.1:\phs          &  \nodata          \\
\cite{Bakos:71}\dotfill      &   4421\tablenotemark{b}  &   $-$0.04\phs          &  \nodata          \\
\cite{Glebocki:72}\dotfill   &  (4828)                  &   $-$0.21~$\pm$~0.25\phs     &  3.3              \\
\cite{Gustaffson:74}\dotfill &  (4630)                  &   +0.04~$\pm$~0.15\phs     &  3.1              \\
\cite{Campbell:78}\dotfill   &   4840                   &   +0.02~$\pm$~0.08\phs     &  \nodata          \\
\cite{Lambert:81}\dotfill    &   \phm{\tablenotemark{d}}5091~$\pm$~100\tablenotemark{d}\phn  &   $-$0.05~$\pm$~0.18\phs  &  3.57~$\pm$~0.46  \\
\cite{Gratton:82}\dotfill    &   4825~$\pm$~60\phn\phn  &   $-$0.04~$\pm$~0.14\phs     &  2.77~$\pm$~0.15  \\
\cite{Kjaergaard:82}\dotfill &  (4790)                  &   +0.04\phs          &  3.1              \\
\cite{Gratton:85}\dotfill    &   \nodata                &   $-$0.06~$\pm$~0.12\phs     &  2.77             \\
\cite{Brown:89}\dotfill      &  (4720)                  &   $-$0.04\phs          &  3.1              \\
\cite{McWilliam:90}\dotfill  &  (4770)                  &    0.00~$\pm$~0.11     &  3.27~$\pm$~0.40  \\
\cite{Luck:95}\dotfill       &  (4650~$\pm$~100)\phn    &   $-$0.02~$\pm$~0.10\phs     &  2.35~$\pm$~0.25  \\
\cite{Mishenina:95}\dotfill  &  (4810~$\pm$~100)\phn    &   $-$0.02~$\pm$~0.10\phs     &  3.00~$\pm$~0.30  \\
\cite{Soubiran:98}\dotfill   &   4769~$\pm$~86\phn\phn  &   $-$0.01~$\pm$~0.16\phs     &  2.98~$\pm$~0.28  \\
\cite{Gray:03}\dotfill       &   4761~$\pm$~80\phn\phn  &   +0.07~$\pm$~0.08\phs     &  3.21             \\
\cite{Santos:04}\dotfill     &   4916~$\pm$~70\phn\phn  &   +0.16~$\pm$~0.08\phs     &  3.36~$\pm$~0.21  \\
\cite{Franchini:04}\dotfill  &   \nodata                &   $-$0.066~$\pm$~0.034\phs   &  \nodata          \\
\cite{Fuhrmann:04}\dotfill   &   4888~$\pm$~80\phn\phn  &   +0.18~$\pm$~0.08\phs     &  3.33~$\pm$~0.10  \\
\cite{Affer:05}\dotfill      &   4935~$\pm$~139\phn     &   +0.14~$\pm$~0.19\phs     &  3.63~$\pm$~0.38  \\
\cite{Luck:05}\dotfill       &   5015~$\pm$~100\phn     &   +0.26~$\pm$~0.11\phs     &  3.49~$\pm$~0.10  \\
This paper\dotfill           &   4800~$\pm$~100\phn     &   \nodata        &  3.1~$\pm$~0.2          \\
\enddata
\tablenotetext{a}{Temperature estimates given in parentheses are
listed for completeness, but are photometric rather than
spectroscopic, and are not considered further.}
\tablenotetext{b}{Although these values are listed as effective
temperatures in the catalog by \cite{Cayrel:01}, they are actually
excitation temperatures. We do not use them here.}
\tablenotetext{c}{The original value reported is +0.27. However,
examination of the iron abundances derived for 12 other stars in this
study indicates the [Fe/H] values are systematically overestimated by
approximately 0.22 dex. Correcting for this offset brings the estimate
for \gamcep\ more in line with the rest of the determinations. We
adopt the revised value here.}
\tablenotetext{d}{The hotter temperature derived in this study is a
consequence of the use of old values of the oscillator strengths
\citep[see][]{Mishenina:95}. We have elected not to use it here.}
\tablecomments{When not reported in the original publications, typical
uncertainties for [Fe/H] have been assumed to be 0.1 dex \citep[0.25
dex for][]{Spinrad:69}, and uncertainties in the effective
temperatures have been assumed to be 100 K.}
\end{deluxetable}


\begin{deluxetable}{lc}
\tablewidth{0pc}
\tablecaption{Photometric estimates of the effective temperature of
\gamcep.\label{tab:photometry}}
\tablehead{\colhead{~~~~~~~~~Photometric system and index~~~~~~~~~} & \colhead{$T_{\rm eff}$ (K)\tablenotemark{a}}}
\startdata
Johnson ($B-V$)\dotfill                 &   4756~$\pm$~53\phn   \\
Str\"omgren ($b-y$)\dotfill             &   4811~$\pm$~76\phn   \\
Vilnius ($Y-V$)\dotfill                 &   4753~$\pm$~79\phn   \\
Vilnius ($V-S$)\dotfill                 &   4741~$\pm$~70\phn   \\
Geneva ($B2-V1$)\dotfill                &   4772~$\pm$~51\phn   \\
Geneva ($B2-G$)\dotfill                 &   4746~$\pm$~44\phn   \\
Geneva ($t\equiv [B2-G]-0.39[B1-B2]$)\dotfill   &   4729~$\pm$~49\phn   \\
Johnson-Cousins ($V-R_C$)\dotfill       &   4696~$\pm$~73\phn   \\
Johnson-Cousins ($V-I_C$)\dotfill       &   4783~$\pm$~52\phn   \\
Cousins ($R_C-I_C$)\dotfill             &   4893~$\pm$~93\phn   \\
DDO $C(42-45)$\dotfill                  &   4672~$\pm$~63\phn   \\
DDO $C(42-48)$\dotfill                  &   4729~$\pm$~54\phn   \\
2MASS ($V-J$)\tablenotemark{b}\dotfill  &   5032~$\pm$~370   \\
2MASS ($V-H$)\tablenotemark{b}\dotfill  &   4972~$\pm$~196   \\
2MASS ($V-K$)\tablenotemark{b}\dotfill  &   4886~$\pm$~209   \\
Tycho ($B_T-V_T$)\dotfill               &   4749~$\pm$~83\phn   \\
Tycho-2MASS ($V_T-K$)\tablenotemark{b}\dotfill  &   4876~$\pm$~194   \\
\enddata
\tablenotetext{a}{Based on the color/temperature calibrations by
\cite{Ramirez:05} for giants, adopting [Fe/H] = $+0.01 \pm 0.05$ and
no reddening (see \S\ref{sec:properties}).}
\tablenotetext{b}{Due to the brightness of \gamcep\ the star was
saturated in the 2MASS measurements and yielded a large photometric
error. This is reflected in the large temperature uncertainty.}

\end{deluxetable}

\clearpage

\clearpage

\clearpage

\clearpage

\clearpage

\clearpage

\clearpage

\clearpage

\clearpage

\clearpage

\clearpage

\clearpage

\clearpage

\clearpage

\appendix

\section{Zero-point corrections to the radial velocities of \gamcep\ from
the literature}
\label{sec:appendix}

The historical sources containing radial velocity measurements of
\gamcep\ typically include other stars observed either as standards or
for other purposes.  The likelihood that many of those stars have been
observed multiple times at the CfA is fairly high given that the
spectroscopic database at CfA contains tens of thousands of stars and
about a quarter of a million spectra to date. This common ground
enables us to place the measurements of each of the sources on the CfA
velocity system. In each case we selected all stars with no obvious
signs of velocity variation that have been observed at least 3 times
at the CfA. Radial velocities were derived from the available CfA
spectra in the same way as those for \gamcep, by cross-correlation
using synthetic templates based on model atmospheres by R.\ L.\ Kurucz
\citep[see][]{Nordstrom:94, Latham:02}. The optimal template for each
star was determined from grids of cross-correlations against a large
number of synthetic spectra over broad ranges in the template
parameters (mainly the effective temperature and rotational velocity),
in the manner described by \cite{Torres:02}. Solar metallicity was
assumed throughout. Many of the stars are giants but there are some
dwarfs as well, so the optimal surface gravity for the template in
each case was determined by repeating the procedure above for a range
of values of $\log g$, and selecting the one giving the highest
correlation averaged over all exposures of the star. The radial
velocities derived with these templates were then compared with those
from each literature source.

Some of these sources have relatively few stars that can be used as
standards, and rejecting objects that have not been observed at CfA
leads to the loss of potentially useful comparison stars in some
cases, which can compromise the determination of the offset.  Those
stars can still be used so long as they are included in another of the
data sets, which then provides the link to the CfA system.  Thus,
instead of separately comparing each source with CfA to determine the
corresponding velocity offset, as might commonly be done, we have
followed a procedure by which we determine the velocity offsets of all
sources simultaneously by minimizing the scatter of the velocities for
all standard stars taken together. In this way any star that is
included in at least two of the data sets (whether or not one of them
is CfA) can be used to strengthen the solution. The quantity we seek
to minimize is
\begin{equation}
\chi^2 = \sum_{i=1}^{N_{\rm sets}} \sum_{j=1}^{N_{i,\rm stars}}
\sum_{k=1}^{N_{ij,\rm obs}} \left({RV_{ijk} - \overline{RV_j} \over
\sigma_{ijk}}\right)^2~,
\end{equation}
 where the sums are performed over all data sets ($i = 1, \ldots, N_{\rm
sets}$), all stars in each data set ($j = 1, \ldots, N_{i,\rm star}$),
and all observations of each star ($k = 1, \ldots, N_{ij,\rm obs}$). The
quantity $\sigma_{ijk}$ represents the uncertainty of each
observation. The mean radial velocity for each star,
$\overline{RV_j}$, is a function the adjustable parameters (offsets
$\Delta_i$) given by
\begin{equation}
\overline{RV_j} = 
\sum_{i=1}^{N_{\rm sets}} \sum_{k=1}^{N_{ij,\rm obs}} \left(RV_{ijk} +
\Delta_i\right)/ \sum_{i=1}^{N_{\rm sets}} N_{ij,\rm obs}~,
\end{equation}
and changes as the iterations proceed.  Since the offsets are computed
relative to CfA (defined here as the first data set), $\Delta_1 \equiv
0$.

Table~\ref{tab:offsets} presents the results for each data set from
our least-squares solution. We list the derived offset along with its
uncertainty, the number of standard stars in each group, the number of
observations of \gamcep, and the interval of those observations. With
a few exceptions the total number of standard star observations used
in each data set is typically a few dozen, while the overall number of
CfA observations used for those same standards is $\sim$3300. The
offsets were added with their corresponding sign to the individual
velocities of \gamcep\ in each data set to place them on the CfA
system.


\begin{deluxetable}{lcccc}[!b]
\tablewidth{0pc} 

\tablecaption{Radial velocity offsets applied to the literature
sources containing measurements of \gamcep, to bring them onto the CfA
system.\label{tab:offsets}}

\tablehead{\colhead{} & \colhead{Offset $\Delta$} & \colhead{Standard}
& \colhead{RVs for} & \colhead{Time span} \\
\colhead{~~~~~~~~~~~~~~~Source~~~~~~~~~~~~~~~} & \colhead{(\kms)} &
\colhead{stars} & \colhead{\gamcep} & \colhead{(yr)}}
\startdata 
\cite{Frost:03}\dotfill      & $-$1.33~$\pm$~0.47 & 12 & 3 & 1902.8--1903.3 \\ 
\cite{Belopolsky:04}\dotfill & $-$0.48~$\pm$~0.93 & \phn5 & 4 & 1903.7 \\ 
\cite{Slipher:05}\dotfill    & $-$0.97~$\pm$~0.58 & \phn9 & 3 & 1905.8 \\ 
\cite{Kustner:08}\dotfill    & $-$1.38~$\pm$~0.70 & 12 & 5 & 1905.9--1907.8 \\ 
\cite{Abt:73}\dotfill        & $+$0.46~$\pm$~0.23 & 14 & 3 & 1916.0--1917.8 \\ 
\cite{Harper:34}\dotfill     & $+$1.82~$\pm$~0.22 & 21 & 1 & 1921.9 \\ 
\cite{Boulon:57}\dotfill     & $-$1.50~$\pm$~0.28 & \phn6 & 1 & 1955.0 \\ 
\cite{Snowden:05}\dotfill    & $+$0.75~$\pm$~0.16 & 14 & 13\phn & 1972.5--1974.7 \\ 
\cite{Kjaergaard:81}\dotfill & $+$0.11~$\pm$~0.48 & 14 & 1 & 1977.7 \\ 
CfA\dotfill                  & \phs0.00 & 71 & 3 & 2001.5--2004.9 \\ 
\enddata
\end{deluxetable}

\end{document}